\documentclass[]{pasj01}

\Received{2024/06/28}
\Accepted{2024/08/20}
 
\usepackage{xcolor}

\newcommand{\kms}{km s$^{-1}$}

\begin{document} 

\title{The XRISM First Light Observation: \\ Velocity Structure and Thermal Properties of the Supernova Remnant N132D
}

\author{XRISM Collaboration\altaffilmark{1}}
\altaffiltext{1}{\textit{Corresponding Authors: Hiroya Yamaguchi, Hiromasa Suzuki, Frederick Scott Porter, Caroline Kilbourne, Michael Loewenstein, Jacco Vink, and Yuken Ohshiro}}
\email{yamaguchi.hiroya@jaxa.jp}


\author{Marc Audard\altaffilmark{2}}
\author{Hisamitsu Awaki\altaffilmark{3}}
\author{Ralf Ballhausen\altaffilmark{4,5,6}}
\author{Aya Bamba\altaffilmark{7}}
\author{Ehud Behar\altaffilmark{8}}
\author{Rozenn Boissay-Malaquin\altaffilmark{9,5,6}}
\author{Laura Brenneman\altaffilmark{10}}
\author{Gregory V.\ Brown\altaffilmark{11}}
\author{Lia Corrales\altaffilmark{12}}
% ...

\altaffiltext{2}{Department of Astronomy, University of Geneva, Versoix CH-1290, Switzerland} 
\altaffiltext{3}{Department of Physics, Ehime University, Ehime 790-8577, Japan} 
\altaffiltext{4}{Department of Astronomy, University of Maryland, College Park, MD 20742, USA} 
\altaffiltext{5}{NASA / Goddard Space Flight Center, Greenbelt, MD 20771, USA} 
\altaffiltext{6}{Center for Research and Exploration in Space Science and Technology, NASA / GSFC (CRESST II), Greenbelt, MD 20771, USA} 
\altaffiltext{7}{Department of Physics, University of Tokyo, Tokyo 113-0033, Japan} 
\altaffiltext{8}{Department of Physics, Technion, Technion City, Haifa 3200003, Israel}
\altaffiltext{9}{Center for Space Science and Technology, University of Maryland, Baltimore County (UMBC), Baltimore, MD 21250, USA}
\altaffiltext{10}{Center for Astrophysics | Harvard-Smithsonian, MA 02138, USA} 
\altaffiltext{11}{Lawrence Livermore National Laboratory, CA 94550, USA} 
\altaffiltext{12}{Department of Astronomy, University of Michigan, MI 48109, USA} 
% ...


\maketitle

\begin{abstract}
We present an initial analysis of the X-Ray Imaging and Spectroscopy Mission (XRISM) first-light observation of the supernova remnant (SNR) N132D in the Large Magellanic Cloud. 
The Resolve microcalorimeter has obtained the first high-resolution spectrum in the 1.6--10\,keV band, which contains K-shell emission lines of Si, S, Ar, Ca, and Fe.  
We find that the Si and S lines are relatively narrow, with a broadening represented by a Gaussian-like velocity dispersion of $\sigma_v \sim 450$\,km\,s$^{-1}$. 
The Fe He$\alpha$ lines are, on the other hand, substantially broadened with $\sigma_v \sim 1670$\,km\,s$^{-1}$. 
This broadening can be explained by a combination of the thermal Doppler effect due to the high ion temperature and the kinematic Doppler effect due to the SNR expansion. Assuming that the Fe He$\alpha$ emission originates predominantly from the supernova ejecta, we estimate the reverse shock velocity at the time when the bulk of the Fe ejecta were shock heated to be $-1000 \lesssim V_{\rm rs}~[{\rm km\,s}^{-1}] \lesssim 3300$ (in the observer frame).  
We also find that Fe Ly$\alpha$ emission is redshifted with a bulk velocity of $\sim 890$\,km\,s$^{-1}$, substantially larger than the radial velocity of the local interstellar medium surrounding N132D.  These results demonstrate that high-resolution X-ray spectroscopy is capable of providing constraints on the evolutionary stage, geometry, and velocity distribution of SNRs. 
\end{abstract}


\section{Introduction}

Supernova remnants (SNRs) play a key role in the process of feedback within galaxies. They inject large amounts of kinetic energy into the interstellar medium (ISM), driving shock waves of thousands of kilometers per second, which heat interstellar gas and dust. 
Their ejecta enrich the ISM with freshly synthesized heavy elements, contributing to the chemical evolution of galaxies. 
Observations of SNRs provide key information about these processes, through their elemental abundances, morphology, velocity distribution, and thermal properties of the shock-heated materials, providing insight into the pre-explosion evolution and explosion mechanism of their progenitors.

The SNR N132D, located in the bar of the Large Magellanic Cloud (LMC), is one of the most studied objects of its class at virtually every wavelength: e.g., radio \citep{dickel95,sano20}, infrared \citep{williams06,rho23}, and gamma-ray \citep{acero16,ackermann16,hess21}. 
Its periphery has an elliptical shape of $\sim 110'' \times 80''$ along the major and minor axes, respectively, which corresponds to $\sim$\,27\,pc $\times$ $\sim$\,19\,pc at the distance to the LMC of 50\,kpc \citep{pietrzynski13}. 
N132D is thought to be in a transitional phase from young to middle-aged (e.g., \cite{bamba18}). Therefore, both swept-up ISM and ejecta contribute to its electromagnetic spectrum.

First identified as an SNR in the radio band \citep{westerlund66}, N132D was categorized as an ``oxygen-rich'' SNR by optical observations (e.g., \cite{danziger76,lasker78,lasker80,dopita84,sutherland95,morse95}). 
Observations with the Hubble Space Telescope (HST) revealed strong emission of the C/Ne-burning products (i.e., O, Ne, Mg) with little emission of the O-burning products (i.e., Si, S), leading to an interpretation of a Type Ib supernova origin \citep{blair00}. 
A survey of the optical emitting ejecta in the [O\,{\footnotesize III}] band revealed that the O-rich ejecta form a toroidal structure with a diameter of $\sim$\,9\,pc, inclined at an angle of 28\,deg to the line of sight \citep{law20}, consistent with earlier work of \citet{vogt11}. 
Assuming homologous expansion from the center of the O-rich ejecta, \citet{law20} derived an average expansion velocity of 1745\,\kms. 
It is worth noting that a toroidal structure of fast-moving ejecta is found in other core-collapse SNRs, such as Cas~A (e.g., \cite{milisavljevic13}), suggesting that the explosion of their progenitors was similarly asymmetric. More recently, the proper motion of the ballistic O-rich ejecta has been measured using multiple epochs of HST data, implying an age of 2770 $\pm$ 500\,yr \citep{banovetz23}, consistent with the age derived from the kinematic reconstruction of \citet{law20}.

The first X-ray detection of N132D was made using the Einstein Observatory \citep{long79,mathewson83}, where the luminosity in the 0.5--3.0\,keV band was reported to be $\sim 4 \times 10^{37}$\,ergs\,s$^{-1}$.
\citet{hughes87} investigated its morphology in detail, and suggested that the SNR had evolved within a cavity in the ISM, likely formed by the pre-explosion stellar wind activity of the progenitor.
Using ASCA, \citet{hughes98} found that the elemental abundances measured for the entire SNR are consistent with the mean LMC values \citep{russell92}, suggesting that the X-ray emission is dominated by the swept-up ISM. 
Later observations using XMM-Newton \citep{behar01} and Chandra \citep{borkowski07} revealed a complex object with ejecta emission mostly coming from the central portion of the remnant, where the X-ray emitting O-rich ejecta have a similar large-scale distribution as the optical emitting ejecta. 
Furthermore, these observations revealed that the Fe K emission, first detected by BeppoSAX \citep{favata97}, exhibits a centrally concentrated morphology, suggesting an ejecta origin.  
A more recent study by \citet{sharda20}, who used the Chandra data, revealed that the Fe K emission is distributed throughout the interior of the southern half of the remnant. 
Suzaku and NuSTAR revealed the presence of Fe Ly$\alpha$ emission in addition to the Fe He$\alpha$ emission \citep{bamba18}, indicating that the Fe ejecta are hot ($kT_{\rm e} \gtrsim 5$\,keV) and highly ionized.

High-resolution X-ray spectroscopy of this SNR has been limited to the soft X-ray band ($< 2$\,keV). 
Using the crystal spectrometer on board the Einstein Observatory, \citet{hwang93} measured the intensities of the forbidden and resonance lines of Ne\,{\footnotesize IX}, which were partially resolved in its spectrum. 
The Reflection Grating Spectrometer (RGS) on board XMM-Newton resolved soft X-ray lines from K-shell ions of C, N, O, Ne, Mg, Si, as well as L-shell ions of S, Ar, Ca and Fe \citep{behar01,suzuki20}. 
In 2016, the Hitomi (ASTRO-H) mission observed N132D, using an X-ray microcalorimeter that could resolve the K-shell lines in the 2--10\,keV band. 
However, it detected only 17 photons in the Fe K band, due to the extremely short exposure ($\sim 3.7$\,ks) caused by attitude control loss during the observation \citep{hitomi18}.

In this paper, we report on the first high-resolution spectroscopy of N132D with high photon statistics in energies above 2\,keV, enabled by the ``first-light'' observation\footnote{https://global.jaxa.jp/press/2024/01/20240105-1\_e.html} of the X-ray Imaging and Spectroscopy Mission (XRISM) \citep{tashiro20}. 
This paper focuses on the velocity structure and thermal properties of the hot plasma that can be constrained by analyzing the well-resolved, strong thermal emission lines detected in the spectrum of the entire SNR. 
The results provide insights into the geometry and dynamical evolution of the SNR, which in turn constrain the nature of the progenitor and its environment. 
Other investigations, including a search for weak emission features and spatially-resolved spectroscopy, are left for future work. 

This paper is organized as follows. Section 2 describes the details of the observations and data reduction. In Section 3, we present data analysis. The implication of the results are discussed in Section 4. Finally, we conclude this study in Section 5. Given that this is one of the first papers on the scientific outcomes of the XRISM mission, we provide detailed descriptions about the gain calibration and background model construction in Appendices. 
The errors quoted in the text and table and error bars given in the figures represent the 1$\sigma$ confidence level, unless otherwise stated.

\section{Observation and Data Reduction}

XRISM was launched aboard the H-IIA Launch Vehicle No.~47 on September 7th, 2023 from JAXA's Tanegashima Space Center. 
The spacecraft was placed into an approximately circular orbit with an inclination of $\sim$\,31\,deg and altitude of $\sim$\,575\,km.
The XRISM scientific payload is comprised of two co-aligned instruments, Resolve \citep{ishisaki22} and Xtend \citep{mori22}, located at the focal plane of two X-ray Mirror Assemblies (XMAs) with the same design. 
The former is an X-ray microcalorimeter, enabling non-dispersive high-resolution spectroscopy in the X-ray band. The latter is a traditional X-ray CCD detector, with a wide field of view of $38' \times 38'$. 
In this paper, we focus on spectroscopy of the Resolve data and use the Xtend data for imaging analysis only. 

The observations of N132D were conducted twice in early December 2023 during its commissioning phase: the first starting at 22:01 UT on December 3 until 00:01 UT on 2023 December 7, and the second starting at 09:53 UT on December 9 until 03:46 UT on December 11. 
The nominal aim point for both observations was (RA, Dec) = (81.$^{\!\circ}$25849, --69.$^{\!\circ}$64122). 
The radial velocity component (toward N132D) of the Earth's orbital motion with respect to the Sun was about $-2.7$\,km\,s$^{-1}$ at the time of the observations.

The data were reduced utilizing the pre-release Build~7 XRISM software and calibration database (CALDB) libraries, representing updates of their Hitomi predecessors\footnote{https://heasarc.gsfc.nasa.gov/docs/hitomi/analysis/}. The data were reprocessed and screened by the automated XRISM processing pipeline version 03.00.011.008 that started on April 16, 2024. 

\subsection{Resolve}

The Resolve observations of N132D were made through a 
$\sim$\,250-$\mu$m thick beryllium window \citep{midooka20} in the closed aperture door, limiting the bandpass to energies above $\sim$\,1.6\,keV.  

The Resolve detector gain and energy assignment require correction of the time dependent gain on-orbit (see Appendix~1 for details). 
A set of $^{55}$Fe radioactive sources on the filter wheel is periodically rotated into the aperture of the instrument to measure the gain.
For the observations of N132D, the gain fiducials were measured every orbit for $\sim 30$ minutes during earth occultation 
yielding 500--600 counts in the Mn K$\alpha$ line complex. 
In total, 72 gain fiducial measurements were conducted. 
Photon energy is assigned to each event using the fiducial gain curves based on the ground calibration and the standard nonlinear energy scale interpolation method \citep{porter16}.
As a result, we achieved an energy resolution of 4.43\,eV (FWHM) and an energy scale error of 0.04\,eV for the N132D observation, confirmed with the fit to the Mn K$\alpha$ spectrum from the $^{55}$Fe radioactive sources (see Appendix~1).
On-orbit measurements using Cr and Cu fluorescent sources and Si instrumental lines give array composite systematic uncertainties in the energy scale of $< 0.2$\,eV in the 5.4--8.0\,keV band and at most 1.3\,eV at the low energy end around 1.75\,keV.

Event-based screening is applied based on that adopted for Hitomi data,
with an updated energy upper limit to the pulse-shape validity (SLOPE\_DIFFER) cut (PI $>$\,22000, 0.5\,eV channels), and including the post-pipeline RISE\_TIME and frame event coincidence screening \citep{kilbourne18}. GTI-filtering is applied to exclude periods of the Earth eclipse and sunlit Earth's limb, SAA passages, and times within 4300\,s from the beginning of a recycling of the 50-mK cooler. The resulting combined effective exposure left after the event screening is 194\,ks. 
We use only Grade (ITYPE) 0 (High-resolution primary) events for the subsequent analysis. 

A redistribution matrix file (RMF) was generated by the {\texttt rslmkrmf} task using the cleaned event file and CALDB based on ground measurements.$\!$\footnote{Following completion of our analysis, we became aware of the presence of non-source low resolution secondary (Ls) events in the cleaned event file that was used to generate the RMF. This issue results in an underestimate of the effective area. To correct it, we scaled the normalizations in the spectral fits by the appropriate factor (the non-Ls fraction). The other spectral parameters are unaffected by this issue.} The following line-spread function components are considered: the Gaussian core, exponential tail to low energy, escape peaks, and Si fluorescence.
An auxiliary response file (ARF) was generated by the {\texttt xaarfgen} task assuming a point-like source at the aim point as an input. We also generated another ARF using a Chandra image in the 0.5--5.0~keV band as an input sky image. These are consistent to better than 4\% across the spectral fitting bandpass. In fact, we confirm no significant difference in results of the following analysis.

\begin{figure*}
 \begin{center}
  \includegraphics[width=13.6cm]{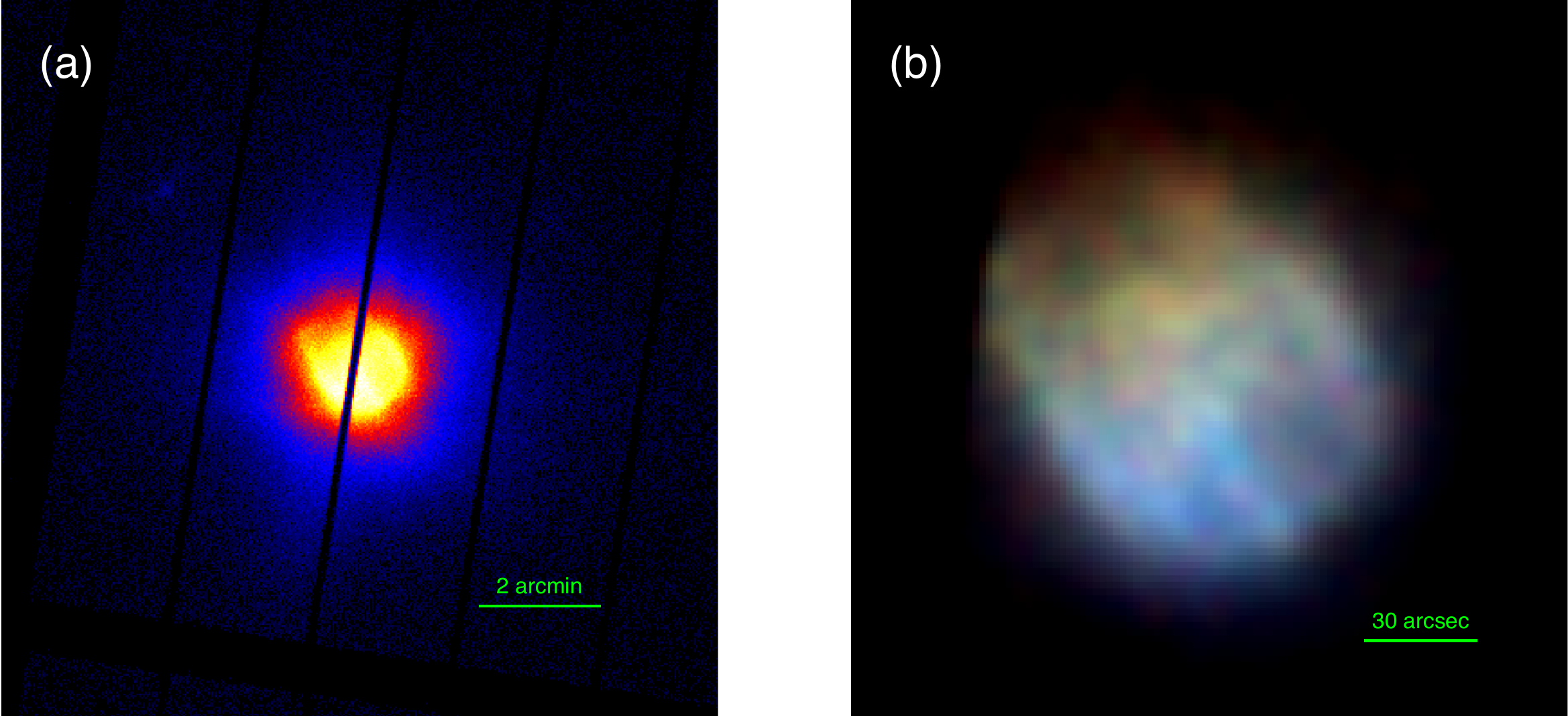} 
 \end{center}
    \caption{(a) Xtend image of N132D obtained with the full-window mode observation conducted from December 3 to 7, 
    where the color corresponds to intensity.
    The `gaps' are due to charge injection rows. \ (b) Xtend image obtained with the 1/8-window mode observation conducted from December 9 to 11. Red, green, and blue correspond to 0.3--0.5\,keV, 0.5--1.75\,keV, and 1.75--10\,keV, respectively.
    }\label{fig:image_xtend}
\end{figure*}

\subsection{Xtend}

The Xtend instrument was operating in the full-window mode during the first observation and in the 1/8-window mode in the second observation, partly for calibration purposes.
After the standard event screening, we obtained effective exposures of 123~ks and 71~ks for the first and second observations, respectively. 
At the time of the observations of N132D, one of the charge-injection rows overlapped with the aim point when the instrument was operating in the full-window mode (Figure\,\ref{fig:image_xtend}a)\footnote{This issue was fixed in 2024 March, so the charge-injection row is no longer overlapping with the aim point.}. 
Therefore, we use only the second observation data for imaging analysis. 
Figure\,\ref{fig:image_xtend}b shows a three-color image of N132D obtained from the second observation: red, green, and blue correspond to 0.3--0.5\,keV, 0.5--1.75\,keV, and 1.75--10\,keV, respectively.

\begin{figure*}
 \begin{center}
  \includegraphics[width=14.6cm]{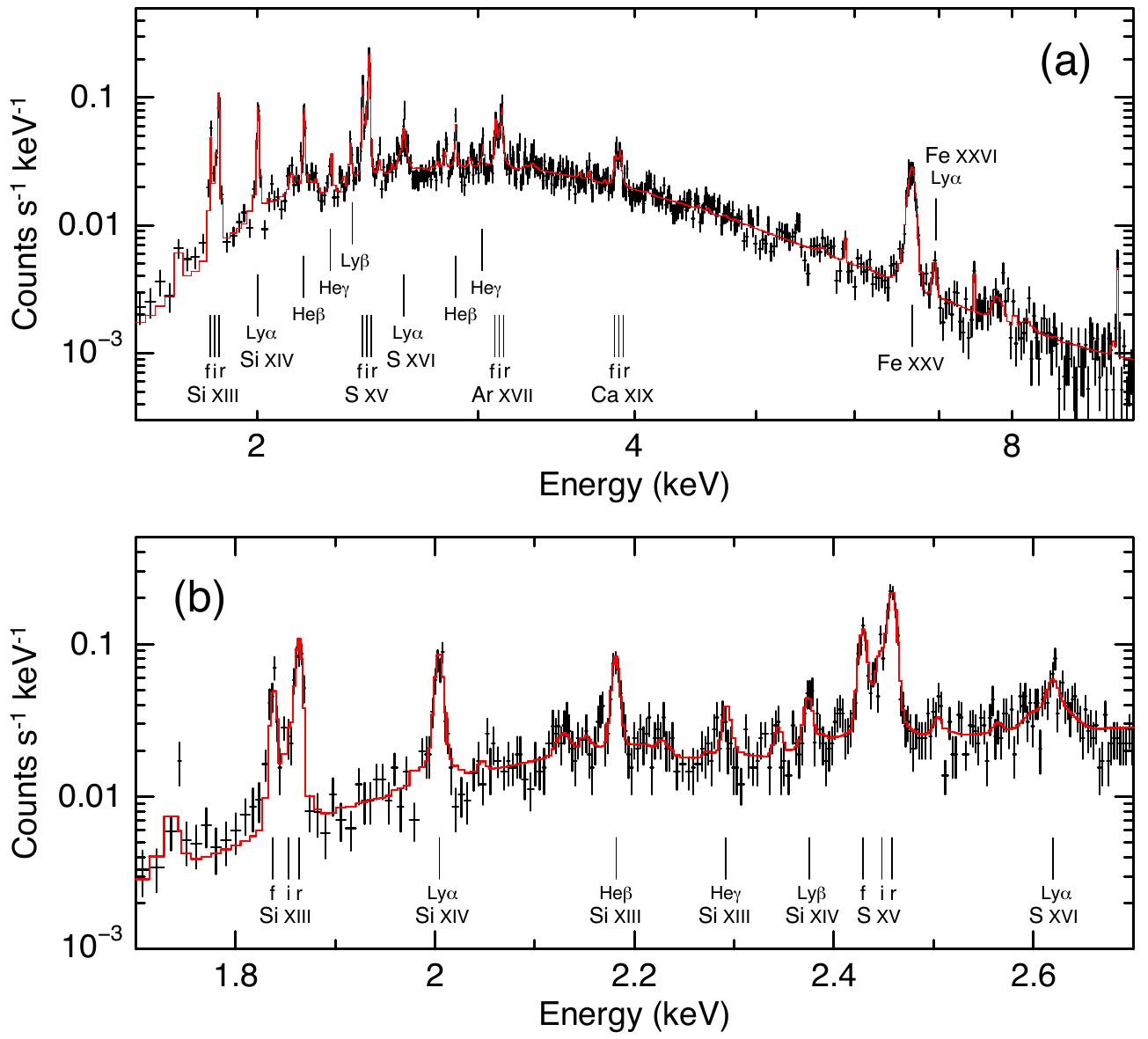} 
 \end{center}
\caption{(a) Resolve spectrum of the SNR N132D in the 1.6--10\,keV band. The red line represents the best-fit Model~B spectrum (whose parameters are given in Table~3). The NXB contribution is also taken into account; the narrow features at 7.5\,keV and 9.7\,keV are Ni-K$\alpha$ and Au L$\alpha$ lines in the NXB, respectively. \ 
(b) Same as panel (a) but magnified in the 1.7--2.7-keV band. 
}
\label{fig:full}
\end{figure*}

\begin{table*}
  \tbl{Emission lines detected in the S K band (2.3--2.6\,keV).}{
  \begin{tabular}{lccccc}
    \hline  
    Transition & $E_{\rm rest}$ (eV)\footnotemark[$*$]  & & $E$ (eV) & $\sigma_E$ (eV) & Norm ($10^{-5}$) \\
    \hline  
    Si\,{\footnotesize XIII} He$\delta$ & 2345.7 & & 2337.4$_{-3.2}^{+4.0}$ & linked to S He$\alpha$ $r$ & 0.73$_{-0.25}^{+0.46}$ \\
    Si\,{\footnotesize XIV} Ly$\beta$ & 2376.4\footnotemark[$\dag$] & & 2376.3$_{-1.8}^{+0.6}$ & 3.8$_{-0.8}^{+1.3}$ & 1.75$_{-0.28}^{+0.43}$ \\
    Si\,{\footnotesize XIII} $n = 9 \rightarrow 1$ (?)\footnotemark[$\ddag$] & 2409.2 & & 2405.1$_{-1.4}^{+2.4}$ & 3.7$_{-1.9}^{+1.5}$ & 0.84$_{-0.21}^{+0.37}$ \\
    S\,{\footnotesize XV} He$\alpha$ ($f$) & 2430.3 & & 2428.8$_{-0.4}^{+0.5}$ & 4.0$_{-0.5}^{+0.4}$ & 5.33$_{-0.37}^{+0.52}$ \\
    S\,{\footnotesize XV} He$\alpha$ ($i$) & 2447.7\footnotemark[$\dag$] & & 2445.9$_{-1.2}^{+0.6}$ & 4.3$_{-0.5}^{+0.2}$ & 3.12 $\pm$ 0.40 \\
    S\,{\footnotesize XV} He$\alpha$ ($r$) & 2460.6 & & 2458.2 $\pm$ 0.3 & 4.0$_{-0.3}^{+0.2}$ & 10.3$_{-0.7}^{+0.4}$ \\
    Si\,{\footnotesize XIII} Ly$\gamma$ & 2506.3\footnotemark[$\dag$] & & 2501.7$_{-2.1}^{+2.2}$ & linked to S He$\alpha$ $r$ & 0.44$_{-0.18}^{+0.26}$ \\
    \hline
    \end{tabular}
    }
\begin{tabnote}
\footnotemark[$*$]Theoretical Rest-frame energies.\\
\footnotemark[$\dag$]Centroid of two lines. \\
\footnotemark[$\ddag$]Details of this emission feature are discussed in a separate paper (XRISM Collaboration, in preparation).
\end{tabnote}
\end{table*}

\section{Analysis}

Figure\,\ref{fig:full} shows the Resolve spectrum in the 1.6--10\,keV band 
extracted from the entire field of view of $3 \times 3$~arcmin$^2$ or 35~pixels. 
The K-shell emission lines of Si, S, Ar, Ca, and Fe are detected with high significance. 
The presence of the Fe Ly$\alpha$ emission, first suggested in the Suzaku study \citep{bamba18}, is also confirmed. 
A remarkable difference is found between the He-like emission lines of the intermediate mass elements (IMEs) and Fe. The former are relatively narrow so the forbidden and resonance lines are well resolved, whereas the latter is significantly broadened.
This is not an instrumental effect, since the spectral resolution (measured in eV) is approximately constant across the 1.6--10\,keV band. 
It is therefore indicative that the IME and Fe emission originate from different plasma components.

In the following subsections, we first look at two narrow band spectra containing the K-shell emission of S (\S3.1) and Fe (\S3.2) separately to investigate the broadening and shift of these line complexes. We then analyze the fullband spectrum with physically realistic models of collisionally ionized plasma (\S3.3).
We apply the optimal binning method of \citet{kaastra16} to the spectrum using the \texttt{ftgrouppha} task in FTOOLS.$\!$\footnote{Throughout the paper, the spectra presented in figures are further rebinned for plotting purposes only. The fitting is performed on the spectrum after the optimal binning.} 
A spectrum of non X-ray background (NXB) is constructed using the method described in Appendix~2 and taken into account in the following analysis. 
The spectral fitting is performed based on the $C$-statistic \citep{cash79}, using the XSPEC software version 12.14.0 \citep{arnaud96}.

\subsection{Sulfur K band}

\begin{figure}
 \begin{center}
  \includegraphics[width=7.8cm]{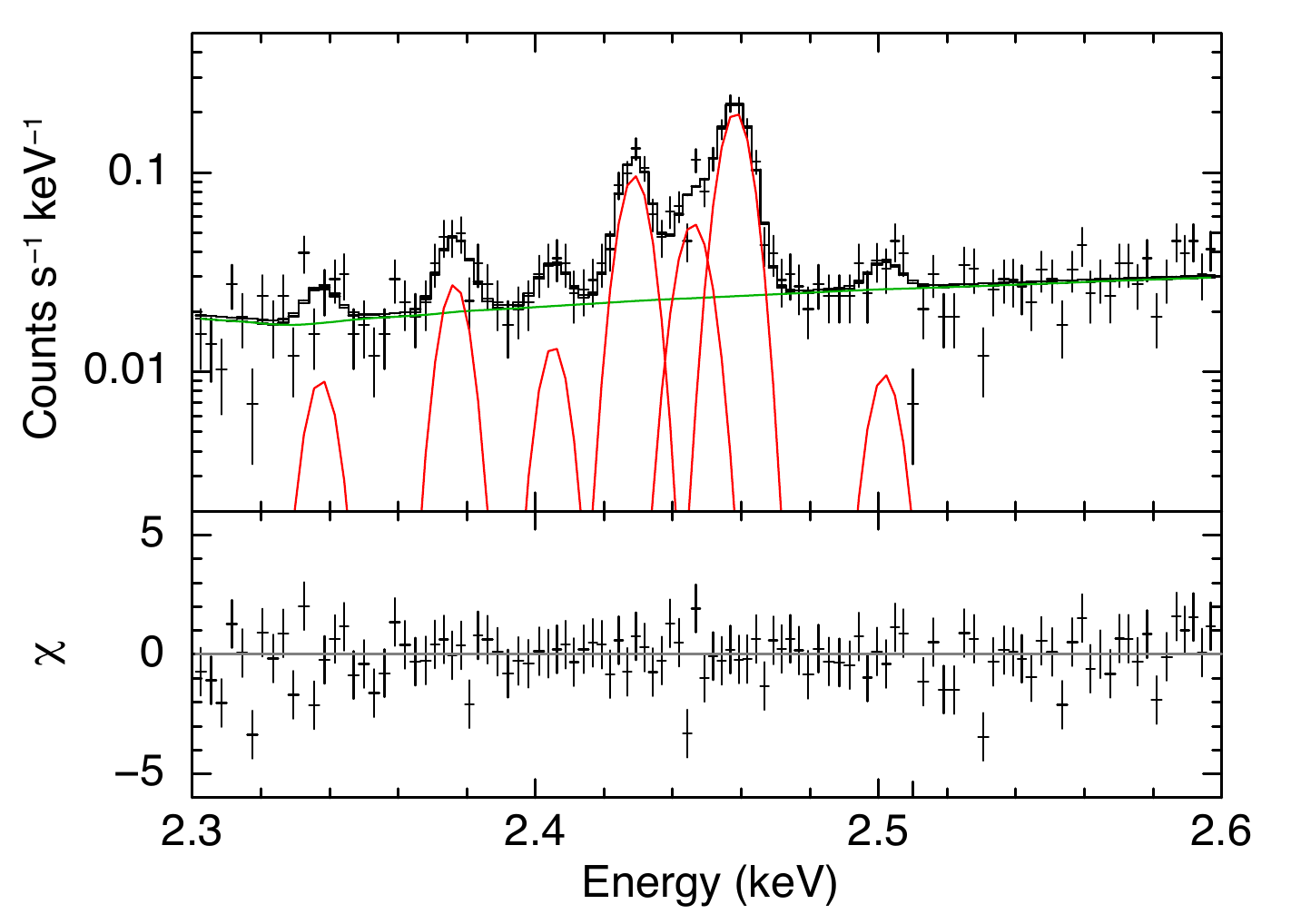} 
 \end{center}
    \caption{The Resolve spectrum in the 2.3--2.6\,keV band, where the S\,{\footnotesize XV} emission is prominent. Red and green are Gaussian functions and the bremsstrahlung continuum component of an ad hoc model, respectively. 
    The NXB contribution is taken into account but is below the displayed flux level.
    }\label{fig:s-band}
\end{figure}

Figure\,\ref{fig:s-band} shows the spectrum around the He-like S emission. There are no spectral features due to the NXB in this band.
To measure the centroid energy and width of the observed emission lines, we fit the spectrum with Gaussian functions and a bremsstrahlung continuum. 
The results are given in Table~1, where the rest-frame energies of the identified lines are also listed. 
We find that both the forbidden ($f$) and resonance ($r$) lines of S XV, whose centroid energies are constrained more stringently than the others, are slightly ($\sim$\,2\,eV) redshifted with respect to their rest frame energies. 
If this shift is purely due to the bulk motion of the plasma, the corresponding velocity is $v \sim 250$\,km\,s$^{-1}$. 
Given the systematic uncertainty in the gain calibration in this energy band (up to 1.3\,eV: Appendix~1), this value is fully consistent with the heliocentric radial velocity of the interstellar gas surrounding N132D, 275$\pm$4\,km\,s$^{-1}$ \citep{vogt11}. 
The 1$\sigma$-width is $\sim$\,4\,eV for these lines, corresponding to a velocity dispersion of $\sigma_v \sim 500$\,km\,s$^{-1}$. 

Interestingly, we detect an unusual emission feature at 2405~eV at a $\sim$\,4$\sigma$ confidence level. 
If real, the most plausible origin would be high-$n$ transitions of Si XIII ($n \sim 9 \rightarrow 1$), suggesting that charge exchange interactions are taking place. 
This interpretation is not unrealistic, given the presence of the dense molecular clouds at the periphery of N132D \citep{williams06,sano20}. 
A more quantitative analysis of the possible charge exchange emission will be presented in a future paper (XRISM Collaboration, in preparation).

\subsection{Iron K band}

\begin{figure}
 \begin{center}
  \includegraphics[width=7.8cm]{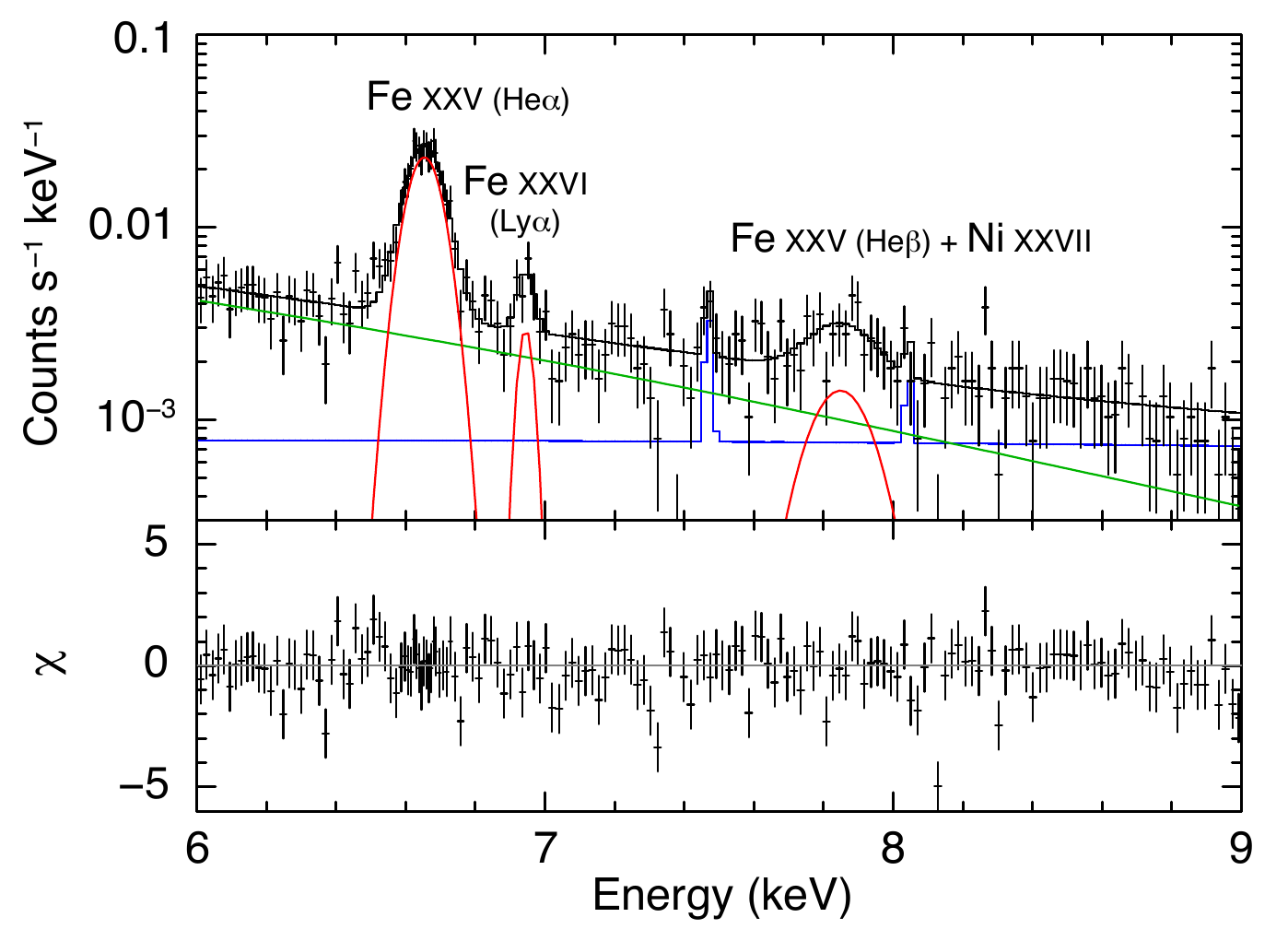} 
 \end{center}
\caption{The Resolve spectrum in the Fe K band. Red and green are the Gaussian functions and bremsstrahlung continuum components of an ad hoc model, respectively. Blue indicates the NXB spectrum. \ 
}
\label{fig:fe-band}
\end{figure}

Figure\,\ref{fig:fe-band} shows the spectrum in the Fe K band. 
The broadened emission at $\sim$\,6.7\,keV predominantly originates from the $n = 2 \rightarrow 1$ transitions of He-like Fe, 
but could also have contributions from multiple features produced by lower-ionization states of Fe. 
The narrower emission feature detected at $\sim$\,6.95\,keV is a mixture of two emission lines, Fe\,{\footnotesize XXVI} Ly$\alpha_1$ and Ly$\alpha_2$. 
In addition, we find a broad feature around 7.9\,keV, possibly a mixture of Fe\,{\footnotesize XXV} $n = 3 \rightarrow 1$ (He$\beta$) emission and Ni\,{\footnotesize XXVII} $n = 2 \rightarrow 1$ (He$\alpha$) emission. 
We note that the contributions of the NXB, indicated by the blue line in Figure\,\ref{fig:fe-band}, are not negligible in this energy band. 
The narrow lines found at 7.5\,keV and 8.0\,keV are due to fluorescence of neutral Ni and Cu, respectively, well reproduced by our NXB model.

Similar to the previous subsection, we fit the spectrum with a phenomenological model consisting of Gaussian functions and a bremsstrahlung continuum, obtaining the results given in Table~2. 
The line parameters are consistent with the previous measurement by \citet{bamba18}. 
We also note that the Fe He$\alpha$ centroid is constrained more stringently than (but consistent with) the Suzaku measurement ($6656 \pm 9$\,eV: \cite{yamaguchi14a}), and significantly more tightly than the XMM-Newton measurement ($6685_{-14}^{+15}$\,eV: \cite{maggi16}). 
We then replace the 6.95-keV Gaussian 
with two \texttt{zgauss} components (whose spectral parameters are the source-frame line energy, redshift, width, and flux) to be able to constrain the bulk velocity shift and broadening of a {\it single} emission line. We fix the line energies to the theoretical rest-frame energies of the Ly$\alpha_1$ and Ly$\alpha_2$ lines (6973\,eV and 6952\,eV, respectively) and the Ly$\alpha_1$/Ly$\alpha_2$ flux ratio to 2 (i.e., statistical weight ratio between the excited states), leaving the redshift and line width as free parameters.
This analysis gives 
a redshift $z = 2.98_{-1.05}^{+1.02} \times 10^{-3}$ and width $\sigma_{\rm E} = 17.4_{-11.9}^{+8.6}$\,eV, 
corresponding to a bulk velocity of $v = 894_{-315}^{+306}$~km\,s$^{-1}$ 
and velocity broadening of $\sigma_v = 749_{-512}^{+370}$~km\,s$^{-1}$.
The bulk velocity is substantially larger than that obtained from the S K band spectrum (and thus larger than the radial velocity of the LMC ISM).

\begin{table}
  \tbl{Emission line complexes in the Fe K band (6--9\,keV).}{
  \begin{tabular}{lccc}
    \hline  
    ~ & $E$ (eV) & $\sigma_E$ (eV) & Norm ($10^{-6}$) \\
    \hline  
    Fe He$\alpha$ etc. & 6654 $\pm$ 3 & 51 $\pm$ 3 & 17.3$_{-0.6}^{+0.9}$ \\
    Fe Ly$\alpha$ & 6945$_{-6}^{+9}$ & 21$_{-5}^{+5}$ & 0.97$_{-0.22}^{+0.30}$ \\
    Fe He$\beta$ + Ni He$\alpha$ & 7850$_{-21}^{+31}$ & 89$_{-19}^{+26}$ & 2.2$_{-0.4}^{+0.5}$ \\
    \hline
    \end{tabular}
    }
\end{table}

\subsection{Fullband spectrum}

\begin{table*}
  \tbl{Best-fit parameters of the spectral fit in the 1.6--10 keV band.}{
  \begin{tabular}{lccccc}
     \hline
     Parameters & & Model A & &  \multicolumn{2}{c}{Model B} \\
     \hline
     ~ & & Low-$T_{\rm e}$ component & & \multicolumn{2}{c}{Low-$T_{\rm e}$ component} \\
     \hline
    $kT_{\rm e}$ (keV) & & 0.79$\pm$ 0.03 & & \multicolumn{2}{c}{0.81$_{-0.04}^{+0.05}$} \\
    $kT_{\rm init}$ (keV) & & 0.01 (fixed) & & \multicolumn{2}{c}{0.01 (fixed)} \\
    $\tau$ ($10^{12}$\,cm$^{-3}$\,s) & & $> 0.46$ & & \multicolumn{2}{c}{0.83$_{-0.31}^{+1.19}$} \\ 
    $z$ ($10^{-4}$) & & 7.4$_{-0.7}^{+0.8}$ & & \multicolumn{2}{c}{7.7 $\pm$ 0.8} \\
    $\sigma_v$ (km\,s$^{-1}$) & & 462 $\pm$ 24 & & \multicolumn{2}{c}{452 $\pm$ 24} \\
    Normalization ($10^{-2}$) & & 7.8 $\pm$ 0.1 & & \multicolumn{2}{c}{7.1 $\pm$ 0.7} \\   
      \hline
     ~ & & High-$T_{\rm e}$ (RP) component & ~ & High-$T_{\rm e}$ component & Very-high-$T_{\rm e}$ component \\
     \hline  
       $kT_{\rm e}$ (keV) & & 2.0 $\pm$ 0.1 & & 1.8 $\pm$ 0.1 & 10 (fixed) \\
       $kT_{\rm init}$ (keV) & & 30 (fixed) & & 0.01 (fixed) & 0.01 (fixed) \\ 
       $\tau$ ($10^{12}$\,cm$^{-3}$\,s) & & 1.0 $\pm$ 0.1 & & $> 0.8$ & 10 (fixed) \\  
       $z$ ($10^{-3}$) & & 1.14$_{-0.31}^{+0.40}$ & & 0.84$_{-0.39}^{+0.32}$ & 2.98  (fixed) \\
       $\sigma_v$ (km\,s$^{-1}$) & & 1700$_{-140}^{+150}$ & & 1670$_{-170}^{+160}$ & 750 (fixed) \\
       Normalization  ($10^{-3}$) & & 9.7 $\pm$ 1.5 & & 9.7 $\pm$ 1.5 & 0.29$_{-0.07}^{+0.06}$ \\       
      \hline
      ~ & & Abundances & & \multicolumn{2}{c}{Abundances\footnotemark[$*$]} \\
      \hline
      Si (solar) & & 0.98$_{-0.07}^{+0.08}$ & & \multicolumn{2}{c}{1.0 $\pm$ 0.1} \\
      S (solar) & & 0.72 $\pm$ 0.05 & & \multicolumn{2}{c}{0.70 $\pm$ 0.04} \\
      Ar (solar) & & 0.85 $\pm$ 0.08 & & \multicolumn{2}{c}{0.79 $\pm$ 0.08} \\
      Ca (solar) & & 0.88$_{-0.12}^{+0.13}$ & & \multicolumn{2}{c}{0.82 $\pm$ 0.12} \\
      Fe, Ni (solar) & & 0.88 $\pm$ 0.08 & & \multicolumn{2}{c}{1.2 $\pm$ 0.1} \\
      \hline
      $C$-stat/dof & & 2435.5/2233 & & \multicolumn{2}{c}{2430.2/2232} \\
      \hline
    \end{tabular}
    }
\footnotemark[$*$]The abundances are linked between the different temperature components.  \\ 
\end{table*}

\begin{figure*}
 \begin{center}
  \includegraphics[width=16.2cm]{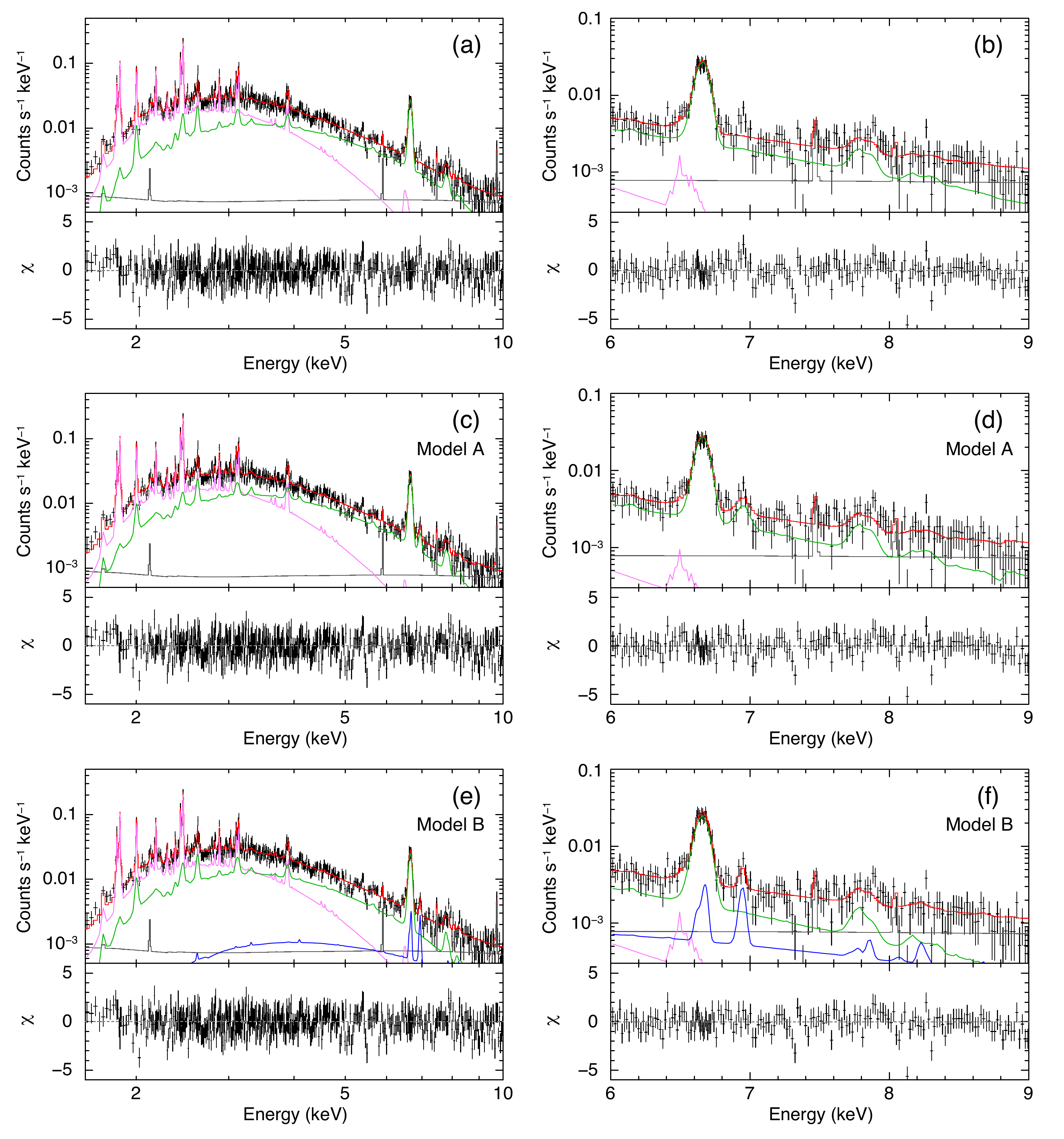} 
 \end{center}
\caption{(a) The Resolve spectrum in the 1.6--10\,keV band, fitted with the two-component ionizing plasma model (red). The contributions of the low-$T_{\rm e}$ and high-$T_{\rm e}$ components are indicated as magenta and green, respectively. Gray indicates the NXB spectrum. \ 
(b) Magnified spectrum in the Fe K band, showing that the Fe Ly$\alpha$ mission is not reproduced by the model. \
(c) and (d) Same as panels (a) and (b), respectively, but fitted with Model~A. \ 
(e) and (f) Same as panels (a) and (b), respectively, but fitted with Model~B. The very-high-$T_{\rm e}$ component is now added, which is indicated as blue. \ 
}
\label{fig:full_fit}
\end{figure*}

We now analyze the Resolve spectrum in 1.6--10\,keV, the energy band with significant signal from the source. 
The analysis in the previous subsections has revealed that different bulk velocities are required to explain the energy shift observed in the S He$\alpha$ and Fe Ly$\alpha$ emission lines. 
Therefore, we start the spectral modeling with two components of a \texttt{bvrnei} model in XSPEC, which reproduces velocity-broadened emission from an optically-thin thermal plasma in non-equilibrium ionization (NEI), either ionizing or recombining. 
The free parameters are the electron temperature ($kT_{\rm e}$), ionization timescale ($\tau = n_{\rm e}t$, where $n_{\rm e}$ and $t$ are the electron density and time elapsed after the abrupt change of temperature, respectively), 
redshift ($z$), velocity dispersion ($\sigma_v$), normalization, and the abundances of Si, S, Ar, Ca, and Fe. 
The abundance of each element is tied between the two components, whereas the other parameters (i.e., $kT_{\rm e}$, $\tau$, $z$, $\sigma_v$) are left independent between the two. 
The Ni abundance is linked to the Fe abundance, and the initial plasma temperature ($kT_{\rm init}$) is fixed to 0.01\,keV (which is equivalent to the assumption that the plasma is ionizing) for both components. 
The foreground absorption is not taken into account in our analysis, because its effect is negligible in this energy band due to the low column density to this SNR ($N_{\rm H} \lesssim 10^{21}$\,cm$^{-2}$) \citep{dickey90,suzuki20}. 
This model yields the best-fit spectrum given in Figure\,\ref{fig:full_fit}a/\ref{fig:full_fit}b. Although the overall spectrum is well reproduced by this model ($C$-stat/dof = 2446.4/2233), it fails to reproduce the observed flux of the Fe Ly$\alpha$ emission. 
We find that the inferred electron temperatures, $kT_{\rm e} \sim 0.83$\,keV and $\sim$\,2.3\,keV (for the low-$T_{\rm e}$ and high-$T_{\rm e}$ components, respectively), are too low to produce a sufficient fraction of H-like Fe ions, if the plasma is ionizing or in equilibrium.

We thus modify the $kT_{\rm init}$ value of one of the {\tt bvrnei} components to 30\,keV for introducing a recombining plasma (hereafter Model~A), similar to the approach taken by \citet{bamba18}. 
The best-fit model spectrum and parameters are given in Figure\,\ref{fig:full_fit}c/\ref{fig:full_fit}d and the ``Model~A'' column of Table~3. 
We find that the flux of both Fe He$\alpha$ complex and Ly$\alpha$ emission are successfully reproduced by the high-$T_{\rm e}$ component.
However, the velocity dispersion of this component ($\sigma_v = 1700_{-140}^{+150}$~km\,s$^{-1}$), which is mainly determined from the width of the Fe He$\alpha$ complex, is significantly larger than that obtained from the Gaussian modeling of the Fe Ly$\alpha$ emission ($\sigma_v = 749_{-512}^{+370}$~km\,s$^{-1}$). 
In fact, the model does not reproduce well the profile of the Ly$\alpha$ line (Figure\,\ref{fig:full_fit}d).

This result leads us to another hypothesis: that different plasma components contribute to the Fe K band spectrum, so that the He$\alpha$ line complex is dominated by a plasma with larger line broadening and the Ly$\alpha$ emission by another plasma with moderate broadening. 
To confirm this possibility, we introduce a third \texttt{bvrnei} component (hereafter very-high-$T_{\rm e}$ component), assuming that all three components are ionizing plasma (hereafter Model~B). 
For the very-high-$T_{\rm e}$ component, the redshift and velocity dispersion are fixed to the values constrained by the \texttt{zgauss} modeling for the Ly$\alpha$ emission (i.e., $z = 2.98 \times 10^{-3}$ and $\sigma_v = 750$~km\,s$^{-1}$). 
Since the electron temperature and ionization age of this component are not well constrained with this complex model, we fix these parameters to $kT_{\rm e}$ = 10\,keV and $\tau = 1\times 10^{13}$\,cm$^{-3}$\,s, and expect that the majority of H-like Fe ions are associated with this component
The best-fit model spectrum and parameters are given in Figure\,\ref{fig:full_fit}e/\ref{fig:full_fit}f and the ``Model~B'' column of Table~3. 
This model yields a slightly better fit than Model~A, especially around the Fe Ly$\alpha$ emission.
We confirm that the line broadening ($\sigma_v$) is significantly larger in the high-$T_{\rm e}$ component (that reproduces the Fe He$\alpha$ complex) than in the very-high-$T_{\rm e}$ component (that reproduces the Fe Ly$\alpha$ emission).
The redshift values are also different between the two components.

In our spectral analysis, the elemental abundances have been tied among the two or three plasma components, since our models are incapable of constraining them independently (i.e., if the abundances of each component are fitted independently, constrained error ranges of several parameters become extremely large). 
The abundances of Si and S are determined mainly by the low-$T_{\rm e}$ component (the magenta curve in Figure\,\ref{fig:full_fit}). Therefore, the actual abundances of Si and S are highly uncertain for the other components. Similarly, the Fe abundance is well constrained only for the high-$T_{\rm e}$ component (the green curve in Figure\,\ref{fig:full_fit}).
Also notable is that the predicted continuum level of the very-high-$T_{\rm e}$ component (the blue curve in Figure\,\ref{fig:full_fit}) is negligibly low compared to the observed continuum level in the whole energy band. Because of this, the observed spectrum can be successfully modeled even if the Fe abundance of this component is set to an extremely high value (e.g., $>$\,10,000\,solar). This implies that the very-high-$T_{\rm e}$ component could be pure-metal, such as in a deep layer of the supernova ejecta.

\subsection{Narrow band image}

\begin{figure*}
 \begin{center}
  \includegraphics[width=15.4cm]{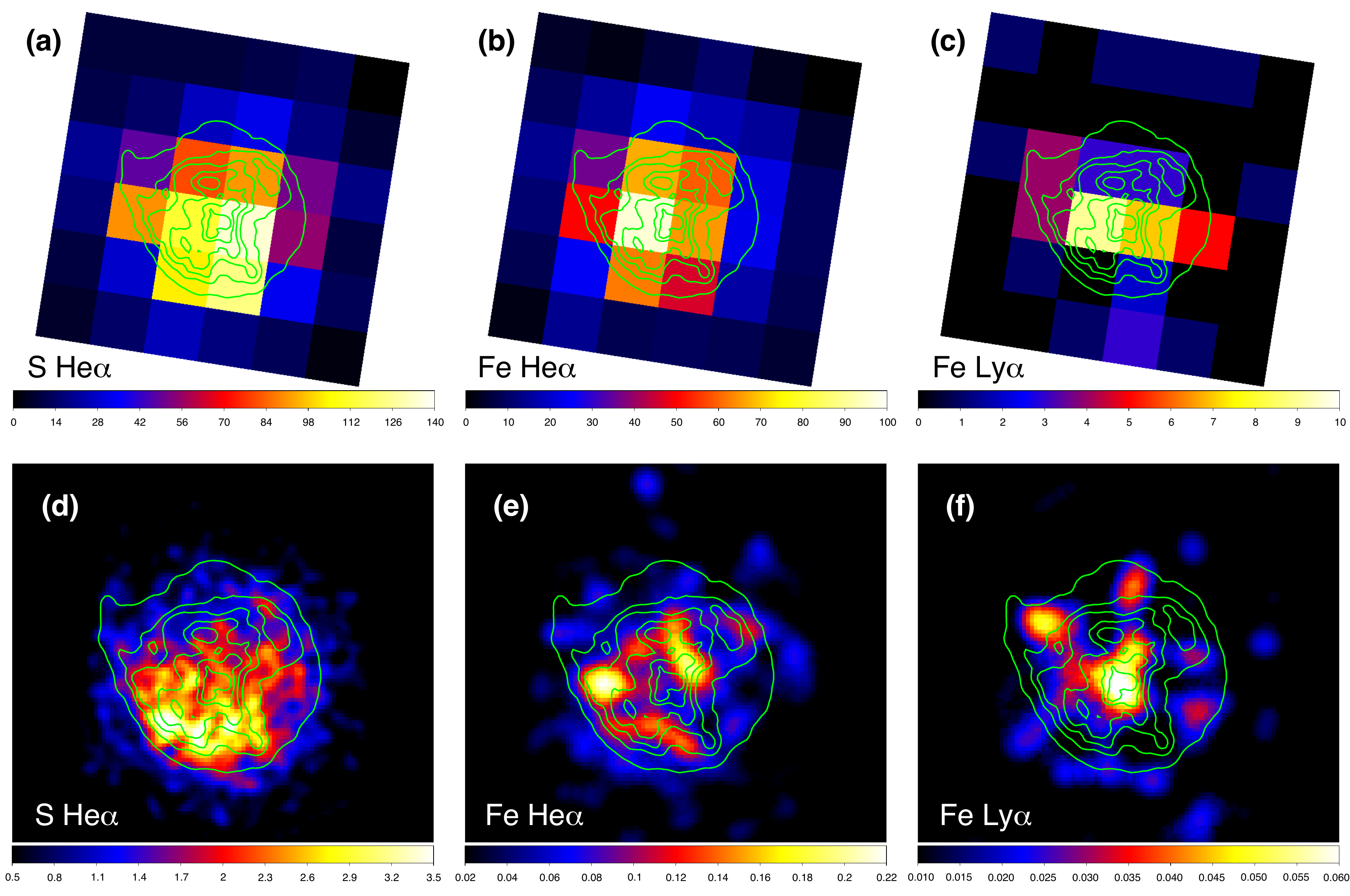} 
 \end{center}
\caption{Top: Raw photon count images of the Resolve in (a)~2.4--2.5\,keV, (b)~6.5--6.8\,keV, and (c)~6.92--6.97\,keV, corresponding to the S He$\alpha$, Fe He$\alpha$, and Fe Ly$\alpha$ emission, respectively. The overplotted contours are the Xtend image in the 0.5--1.75\,keV band. \ 
Bottom: Smoothed photon count image of the Xtend in (d)~2.3--2.6\,keV, (e)~6.5--6.8\,keV, and (f)~6.85--7.05\,keV, respectively.
}\label{fig:image}
\end{figure*}

The result based on Model~B in the previous subsection implies that the Fe He$\alpha$ and Ly$\alpha$ emissions originate from different plasma components. If this hypothesis is true, different spatial distributions of these emissions should be expected. 
We thus generate narrow band images of the Fe He$\alpha$ and Ly$\alpha$ emission as well as the S He$\alpha$ emission to search for morpological differences. The results are shown in Figure\,\ref{fig:image}; the top and bottom panels are the Resolve and Xtend images, respectively.
The distribution of the Fe Ly$\alpha$ emission is localized near the SNR center, whereas the Fe He$\alpha$ emission is more widely distributed. 
This difference is seen by both instruments, although the low statistics in the emission lines makes it difficult to determine their true morphology.
The morphological difference between the S and Fe emission is also confirmed, consistent with previous studies \citep{behar01,borkowski07,sharda20}.

\section{Discussion}

We have performed line-resolved spectroscopy of the thermal emission from N132D, using the XRISM first-light data. 
The Resolve spectrum in the 1.6--10\,keV band can be modeled by two or three components of NEI plasmas with different electron temperatures. 
The K-shell emission lines of Si and S are characterized by an ionizing plasma with the electron temperature of $\sim$\,0.8\,keV. 
The spectrum in the Fe K band (6--9\,keV) is, on the other hand, reproduced by either a one-component recombining plasma (Model~A) or two-component ionizing plasmas (Model~B) with higher temperature. Although the goodness of the fit is comparable between the two models, the different spatial distributions between the Fe He$\alpha$ and Ly$\alpha$ emission, revealed by our imaging analysis (Figure\,\ref{fig:image}), favors Model~B as the more likely scenario. 
The energy shift and broadening of the thermal emission lines have been investigated to constrain the velocity structure of each component. 
Table~4 summarizes the results based on Model~B, where $v_{\rm bulk}$ is the heliocentric radial velocities corrected to the solar system barycentric standard of rest. We discuss the interpretation in the following subsections.

\begin{table}
  \tbl{Summary of the measured bulk velocity and velocity dispersion. \label{tab:broadening}}{
  \begin{tabular}{lccc}
    \hline  
    ~ & Si \& S He$\alpha$\footnotemark[$*$] & Fe He$\alpha$\footnotemark[$*$] & Fe Ly$\alpha$\footnotemark[$*$] \\
    \hline  
    $v_{\rm bulk}$ (km\,s$^{-1}$) & 227 $\pm$ 24 & 249$_{-117}^{+96}$ & 891$_{-315}^{+306}$ \\
    $\sigma_v$ (km\,s$^{-1}$) & 452 $\pm$ 24 & 1670$_{-170}^{+160}$ & 749$_{-512}^{+370}$ \\
     \hline
    \end{tabular}
    }
\begin{tabnote}
\footnotemark[$*$] Reproduced by the low-$T_{\rm e}$, high-$T_{\rm e}$, and very-high-$T_{\rm e}$ components, respectively. For the Fe Ly$\alpha$ emission, the given uncertainties are determined from the \texttt{zgauss} modeling described in Section~3.2.
\end{tabnote}
\end{table}

\subsection{Radial velocity}

We have shown that all the detected emission lines from the IMEs (Si and S) and Fe are redshifted with respect to their rest frame energies. 
The bulk velocity measured from the IME lines and Fe He$\alpha$ complex are consistent with the radial velocity of the interstellar gas surrounding N132D, 275$\pm$4\,km\,s$^{-1}$ \citep{vogt11}. 
On the other hand, the Fe Ly$\alpha$ emission indicates a larger velocity of $\sim$\,900\,km\,s$^{-1}$.
Notably, the Hitomi SXS study of this SNR indicated a similarly high bulk velocity of $\sim$\,1080\,km\,s$^{-1}$ \citep{hitomi18}, but this estimate was obtained using only 17 photons detected in the Fe He$\alpha$ band (not in the Ly$\alpha$ band). Our measurement of the Fe He$\alpha$ bulk velocity, 249$_{-117}^{+96}$\,km\,s$^{-1}$, is lower than the mean value of the Hitomi measurement, but still within its 90\% confidence interval of 330--1780\,km\,s$^{-1}$. 

The large redshift observed in the Fe Ly$\alpha$ emission implies that this emission originates from the Fe-rich ejecta with a highly asymmetric velocity distribution. The ejecta scenario is also supported from the high electron temperature that can be achieved by a high-velocity shock.
The morphology of the Fe Ly$\alpha$ emission is centrally concentrated (Figure\,\ref{fig:image}), suggesting that this hot Fe ejecta component is present only at the far side of the SNR. Theoretically, an ionization timescale of $\tau \gtrsim 10^{12}$\,cm$^{-3}$\,s is required to produce a sufficient fraction of H-like Fe in an ionizing plasma. Therefore, from the estimated SNR age of 2770\,yr \citep{banovetz23}, the electron density of this component is estimated to be $n_{\rm e} = \tau / t_{\rm age} \gtrsim 11$\,cm$^{-3}$. Such a high density implies that the Fe ejecta form a compact knot. 
In fact, the normalization of the very-high-$T_{\rm e}$ component given in Table~3 corresponds to the emitting volume of 
$\sim 7 \times 10^{55}\,(n_{\rm e}/11\,{\rm cm}^{-3})^{-2}$\,cm$^3$, less than 0.1\% of the total SNR volume ($\sim 10^{59}$\,cm$^3$).

A highly asymmetric distribution of Fe ejecta has been observed in other core-collapse SNRs, such as Cas~A \citep{delaney10,hwang12} and G350.1--0.3 \citep{borkowski20,tsuchioka21}. Recent NuSTAR observations of Cas~A and SN\,1987A also revealed that the velocity distribution of radioactive $^{44}$Ti, produced in the same nuclear processes that produce $^{56}$Ni (parent nucleus of Fe), is asymmetric toward one side \citep{grefenstette14,boggs15}, similar to what we observe in N132D. 
Such asymmetry could have been produced by a supernova explosion involving asymmetric effects, such as a standing accretion shock instability \citep{blondin03,janka16}, or an asymmetric interaction between the SNR ejecta and ambient medium. 
For N132D the latter scenario is not unlikely, since the SNR is known to be interacting with dense molecular clouds \citep{sano20}.

\subsection{Origin of the line broadening}

One remarkable finding from our high-resolution spectroscopy is that the Fe He$\alpha$ lines are substantially broadened, whereas the K-shell emission lines from Si and S are only slightly broadened (Table~\ref{tab:broadening}).
There are two plausible causes for the line broadening: (1) thermal Doppler broadening due to high ion temperature or (2) a variation of bulk motion along the line of sight.
Here we simply assume $\sigma_v = \sqrt{ \sigma_{\rm th}^2 + \sigma_{\rm kin}^2 }$, where $\sigma_{\rm th}$ and $\sigma_{\rm kin}$ indicate the broadening due to (1) and (2), respectively.
Both depend on the shock velocity, or more precisely, the upstream bulk velocity in the shock-rest frame, $v_{u,{\rm sh}}$. 
In collisionless shocks, which are generally formed in SNRs, temperature equilibration among different species is not necessarily achieved at the immediate downstream region. 
In the most extreme case, the relation between the shock velocity and downstream temperature, derived from the Rankine-Hugoniot equations, hold independently among different species $i$ as
\begin{equation}
    kT_i = \frac{3}{16} m_i v_{u,{\rm sh}}^2,
    \label{eq:RH}
\end{equation}
where $T_i$ and $m_i$ are the temperature and mass, respectively (e.g., \cite{vink15}). 
Although the process called ``collisionless electron heating'' slightly modifies the temperatures from those predicted by Equation~\ref{eq:RH} (e.g., \cite{ghavamian07,yamaguchi14b}), the effect is not essential in the situations we discuss below. 
The different species then slowly equilibrate to a common temperature via Coulomb collisions in further downstream regions, the timescale of which is discussed later. 
The thermal Doppler broadening is given as 
\begin{equation}
   \sigma_{\rm th} = \sqrt{\frac{kT_i}{m_i}}. 
\end{equation}
Therefore, $\sigma_{\rm th} = (\sqrt{3}/4)\cdot v_{u,{\rm sh}}$ is expected when Equation\,\ref{eq:RH} is strictly achieved.
On the other hand, the downstream bulk velocity in the shock-rest frame is expected as $v_{d,{\rm sh}} = (1/4) \cdot v_{u,{\rm sh}}$,
assuming a compression ratio of 4. 
Therefore, the velocity in the observer frame is calculated as
\begin{equation}
    v_{d,{\rm obs}} = V_{\rm s} + v_{d,{\rm sh}} 
    = V_{\rm s} + \frac{1}{4} v_{u,{\rm sh}},
    \label{eq:v}
\end{equation}
where $V_{\rm s}$ is the shock velocity in the observer frame. 
The line profile depends on the velocity distribution and is not always Gaussian-like. 
As described in Appendix~3, we can approximate as $\sigma_{\rm kin} \approx |v_{d,{\rm obs}}|/2$, when the distributions of the plasma density and velocity are spherically symmetric (e.g., expanding shell or sphere).

\begin{figure}
 \begin{center}
  \includegraphics[width=7.8cm]{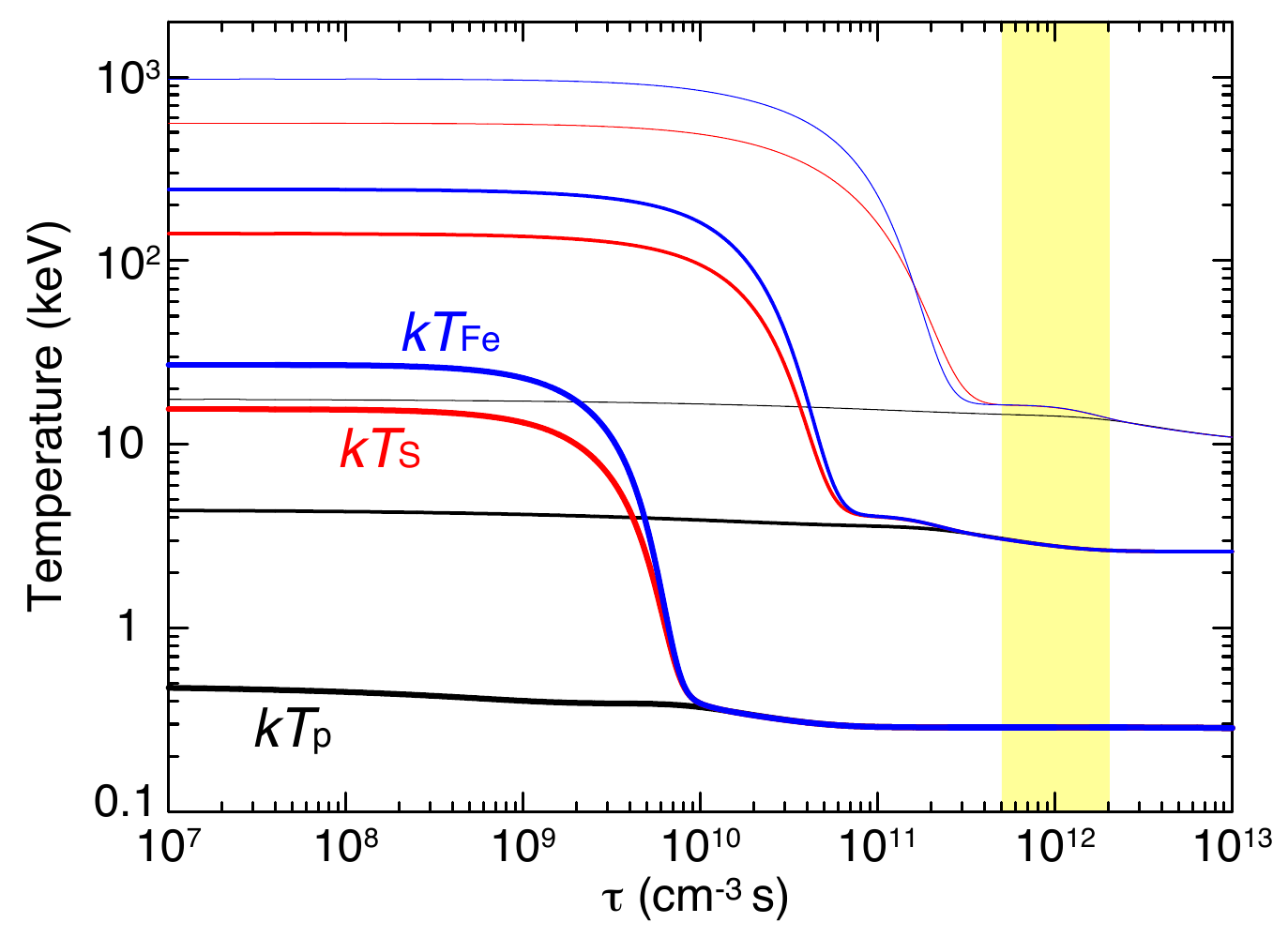} 
 \end{center}
    \caption{Thermal equilibration process among different species via Coulomb collisions in post-shock plasma initially shocked by the blast wave propagating into the ISM. Black, red, and blue indicate temperatures of protons, sulfur, and iron, respectively. 
    From bottom (thick curves) to top (thin curves), the upstream bulk velocities (i.e., blast wave velocity in the observer frame) of 500, 1500, and 3000\,km\,s$^{-1}$ are assumed.
    The elemental abundances measured by \citet{suzuki20} are assumed. 
    The yellow region indicates the range of $\tau$ constrained for the low-$T_{\rm e}$ component.  
    }\label{fig:equil_ISM}
\end{figure}

The previous Chandra study revealed that the K-shell emission from Si and S forms the outermost shell (e.g., \cite{borkowski07,sharda20}), suggesting that the emission predominantly originates from the ISM shocked by the SNR blast wave (although the abundances slightly larger than the mean LMC values of \citet{dopita19} may imply a small contribution of ejecta to the emission).
Figure\,\ref{fig:equil_ISM} shows the thermal equilibration processes due to the Coulomb collisions in the postshock plasma, 
where various shock velocities (that determine the initial temperatures) and the ISM abundances measured by \citet{suzuki20} are assumed. The temporal evolution of $kT_{\rm p}$, $kT_{\rm S}$, and $kT_{\rm Fe}$ are given as a function of $\tau$. 
It is clearly indicated that the ion temperatures are equilibrated with the proton temperature at the timescale constrained for the low-$T_{\rm e}$ component (i.e., $\tau$ = (0.5--2) $\times \, 10^{12}$\,cm$^{-3}$\,s: Table~3). 
Therefore, the thermal Doppler broadening of the S He$\alpha$ lines in the ISM component is calculated to be 
\begin{equation}
    \sigma_{\rm th} = \sqrt{\frac{kT_{\rm S}}{m_{\rm S}}} 
    \approx \sqrt{\frac{kT_{\rm p}}{m_{\rm S}}}
    = \sqrt{\frac{3}{16}\frac{m_{\rm p}}{m_{\rm S}}} \cdot V_{\rm bw},
\end{equation}
where $V_{\rm bw} \ (= - v_{u,{\rm sh}})$ is the blast wave velocity. 
The line broadening due to the bulk expansion is also obtained as 
\begin{equation}
    \sigma_{\rm kin} \sim |v_{d,{\rm obs}}|/2 = \left( V_{\rm bw} - \frac{1}{4}V_{\rm bw} \right) \cdot \frac{1}{2} = \frac{3}{8}V_{\rm bw}, 
\end{equation}
which is much larger than $\sigma_{\rm th}$. 
Therefore, we obtain $\sigma_v = \sqrt{ \sigma_{\rm th}^2 + \sigma_{\rm kin}^2 } \approx (3/8)\cdot V_{\rm bw} \approx 450\,{\rm km}\,{\rm s}^{-1}$, and thus $V_{\rm bw} \approx 1200\,{\rm km}\,{\rm s}^{-1}$.

\begin{figure}
 \begin{center}
  \includegraphics[width=7.8cm]{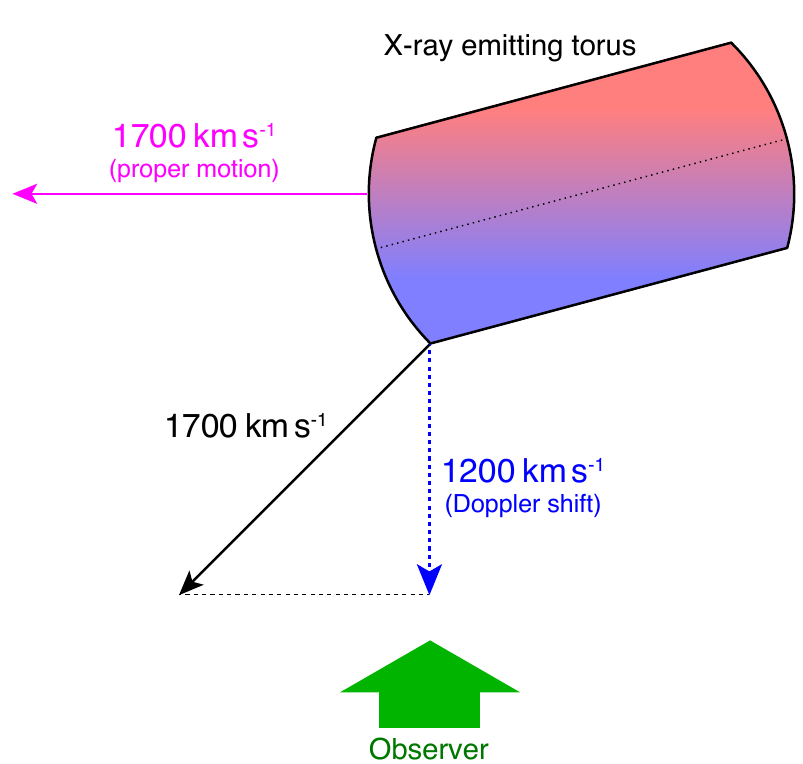} 
 \end{center}
    \caption{Top view of the interpreted geometry and velocity structure of the shocked ISM. The blueshifted and redshifted regions are shown in blue and red, respectively. The primary plane of the torus is indicated by the dotted line, but its inclination angle is arbitrary, as it is not constrained in this work.} 
    \label{fig:schematic}
\end{figure}

Using the proper motion detected with Chandra, \citet{plucinsky24} has measured the blast wave velocity at the bright southern rim to be $1709 \pm 386$~km\,s$^{-1}$, the mean of which is 1.4 times higher than our measurement. 
We note that this method directly measures the current shock velocity, whereas the Doppler broadening reflects the shock velocity in the past. It is, therefore, surprising that a larger velocity is obtained from the former. 
This discrepancy can be explained if the shocked ISM forms a toroidal or ellipsoidal structure with a somewhat low inclination angle as illustrated in Figure\,\ref{fig:schematic}, rather than a spherically symmetric shell. 
This interpretation is supported by the fact that the line broadening is well characterized by a Gaussian profile, because a shell-like geometry with a spherically symmetric velocity distribution leads to a top-flat line profile (see Appendix~3).

As mentioned in Section~1, optical observations of N132D revealed a toroidal geometry of the O-rich ejecta \citep{lasker80,vogt11} with an inclination angle of $\sim$\,28\,deg \citep{law20}. The radius of this torus is $\sim$\,4.5\,pc, smaller than the projected forward shock radius of $\sim$\,10\,pc. 
\citet{vogt11} suggested that the X-ray emitting ISM shell also forms a toroidal structure and that the tori of the ISM and optical emitting ejecta are aligned and thus physically associated with each other.
If the three-dimensional geometry of the SNR is indeed torus-like, the progenitor of N132D must have exploded within a dense CSM with a disk-like geometry. Such CSM distribution is expected if the progenitor is in a binary system, because the pre-explosion stellar wind forms a dense CSM disk on the equatorial plane (e.g., \cite{smith17}). 
A similar scenario is suggested for SN\,1987A (e.g., \cite{podsiadlowski17}), where a dense CSM ring is observed in various wavelengths (e.g., \cite{ravi24} and references therein).

The plasma component responsible for the Fe He$\alpha$ emission has a higher electron temperature and larger line broadening than that responsible for the S He$\alpha$ emission (Table~3). 
If the Fe He$\alpha$ emission originates from the swept-up ISM, this component should have been shock-heated when the blast wave velocity was much higher than it is today. Similar to the previous estimate, we derive $V_{\rm bw} \approx 4450\,{\rm km}\,{\rm s}^{-1}$ in this case, not unreasonable as a shock velocity in the early evolutionary stage of an SNR. 

\begin{figure}
 \begin{center}
  \includegraphics[width=7.8cm]{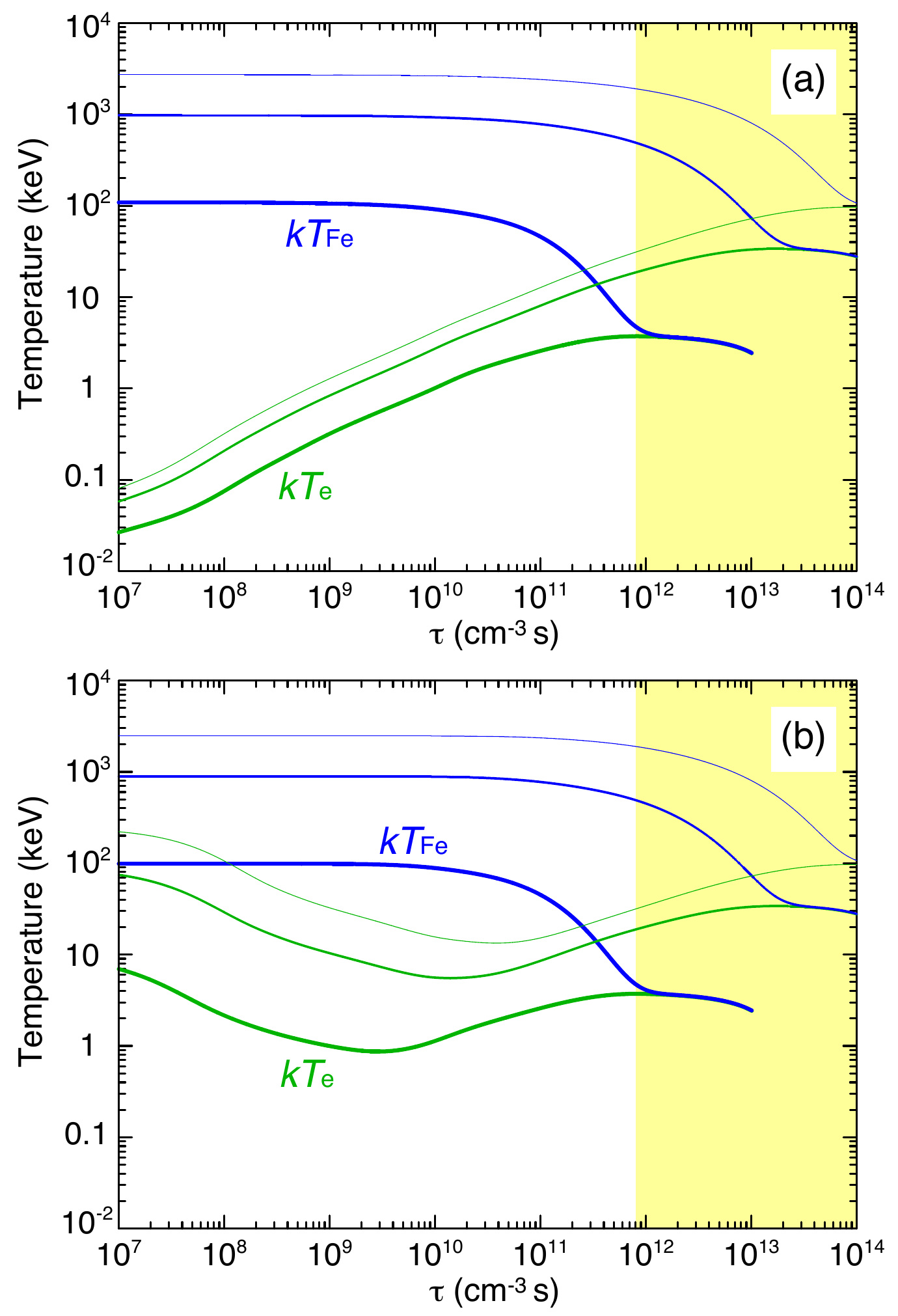} 
 \end{center}
    \caption{Thermal equilibration between Fe temperature (blue) and electron temperature (green) via Coulomb collisions in shocked ejecta with pure-Fe composition. 
    From bottom (thick curves) to top (thin curves), the upstream bulk velocities of 1000, 3000, and 5000\,km\,s$^{-1}$ are assumed. 
    Panels~(a) and (b) assume the initial temperature ratio ($kT_{\rm e}/kT_{\rm Fe}$) of $m_{\rm e}/m_{\rm Fe}$ (i.e., no collisionless electron heating at the reverse shock) and 0.1 (i.e., efficient collisionless electron heating), respectively.
    More details about the calculations are described in Ohshiro et al., in preparation.
    The yellow region indicates the range of $\tau$ constrained for the high-$T_{\rm e}$ component.
    }\label{fig:equil_ej}
\end{figure}

An alternative, more likely possibility is that the SN ejecta contributes significantly to the Fe He$\alpha$ emission, as suggested by some previous work (e.g., \cite{sharda20}). 
In shocked ejecta, especially when it consists purely of heavy elements, the expected thermal evolution properties are distinct from the shocked ISM. 
This is because there are few or no protons or helium nuclei to first equilibrate with, and thus the heavy elements equilibrate with free electrons released from the heavy element atoms themselves. 
This is quantitatively shown in Figure\,\ref{fig:equil_ej}, where the temporal evolution of $kT_{\rm Fe}$ in a pure-Fe plasma is calculated for two cases with different assumptions for the initial conditions: (a) no collisionless electron heating taking place at the reverse shock, or (b) $kT_{\rm e}/kT_{\rm Fe} = 0.1$ is achieved immediately behind the reverse shock due to the efficient collisionless electron heating. 
Unlike the LMC abundance case (Figure\,\ref{fig:equil_ISM}), the Fe ions remain at their initial temperature until $\tau \sim 10^{12}$\,cm$^{-3}$\,s, if the shock velocity is high enough ($\gtrsim 3000$\,km\,s$^{-1}$). 
This conclusion is not affected by the efficiency of the collisionless electron heating.
Therefore, the thermal Doppler broadening is non-negligible in this case.

In general, the reverse shock in an SNR initially moves outward, and then reverses direction moving inward after a few hundred to thousand years, depending on the ambient density structure and other properties (e.g., \cite{truelove99}). 
The upstream fluid velocity in the shock-rest frame, which determines the heating properties, is expressed as 
\begin{equation}
    v_{u,{\rm sh}} = \frac{R_{\rm rs}}{t} - V_{\rm rs},
\end{equation}
where $R_{\rm rs}$ is the reverse shock radius (and thus $R_{\rm rs}/t$ is the free expansion velocity of the outermost unshocked ejecta) and $V_{\rm rs} (= dR_{\rm rs}/dt)$ is the reverse shock velocity in the observer frame (e.g., \cite{vink22}). 
Since Equations~\ref{eq:RH} and \ref{eq:v} hold in the shock-rest frame, 
\begin{equation}
    kT_{\rm Fe} 
    = \frac{3}{16} \, m_{\rm Fe} \left( \frac{R_{\rm rs}}{t} - V_{\rm rs} \right)^2
    \label{eq:kT_Fe}
\end{equation}
and
\begin{equation}
    v_{\rm ej,obs} = \frac{1}{4} \frac{R_{\rm rs}}{t}
    + \frac{3}{4} V_{\rm rs}
\end{equation}
are obtained, where $v_{\rm ej,obs}$ is the bulk velocity of the shocked ejecta in the observer frame.
Assuming that the current Fe temperature is still comparable to the immediate postshock temperature (i.e., Eq.\,\ref{eq:kT_Fe}), we obtain the thermal Doppler broadening to be
\begin{equation}
    \sigma_{\rm th} = \frac{\sqrt{3}}{4} \left( \frac{R_{\rm rs}}{t} - V_{\rm rs} \right).
\end{equation}
From Equations~8 and 9 and an assumption of $\sigma_{\rm kin} \sim |v_{\rm ej,obs}|/2$, the total broadening is obtained as 
\begin{equation}
    \sigma_v 
    = \sqrt{\frac{13}{64} \left( \frac{R_{\rm rs}}{t} \right)^2 - \frac{18}{64}\frac{R_{\rm rs}}{t}V_{\rm rs} + \frac{21}{64} V_{\rm rs}^2}~. 
\end{equation}

\begin{figure}
 \begin{center}
  \includegraphics[width=7.8cm]{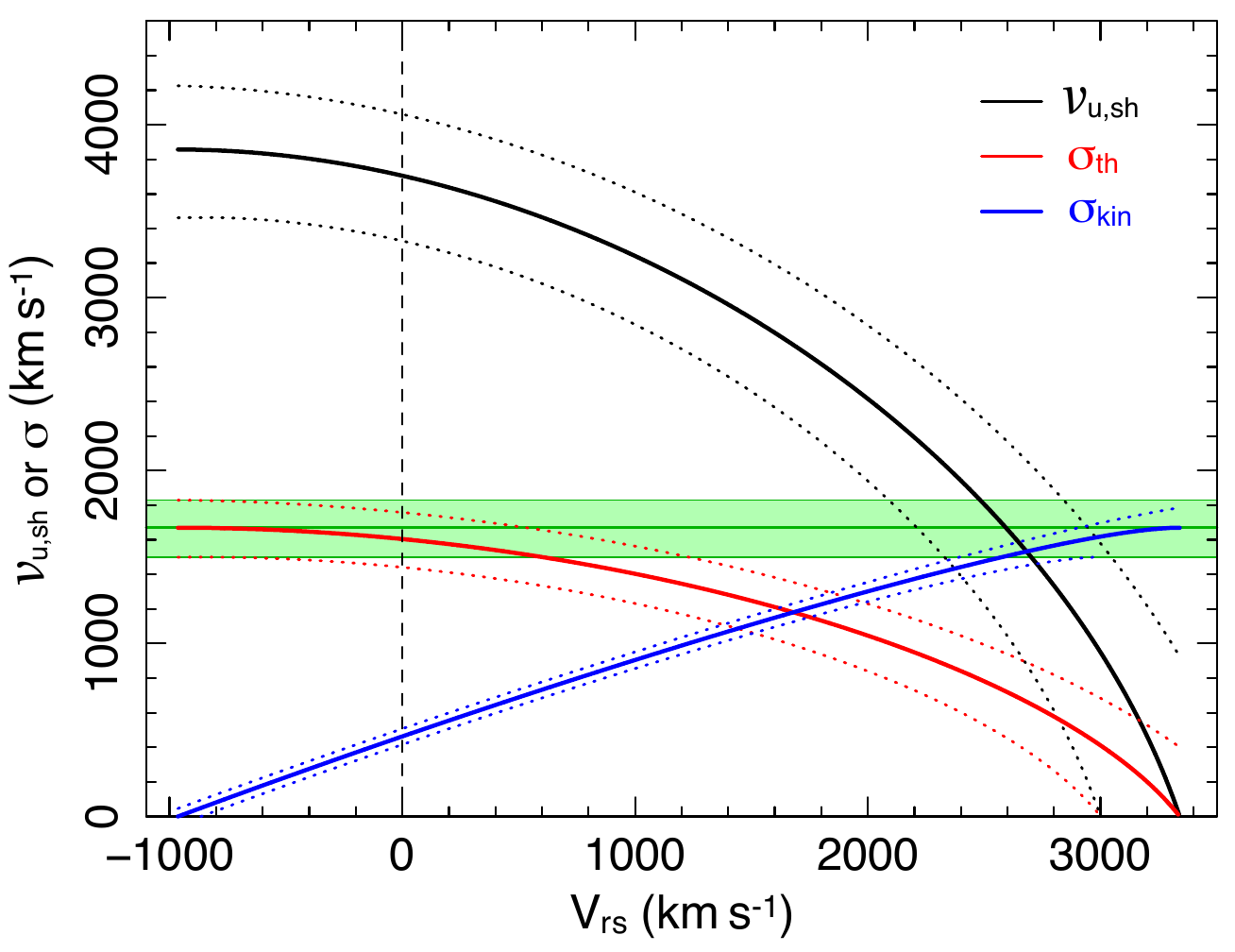} 
 \end{center}
    \caption{Relation between the reverse shock velocity 
    in observer frame ($V_{\rm rs}$) and the upstream bulk velocity 
    in shock-rest frame ($v_{\rm u,sh} = R_{\rm rs}/t - V_{\rm rs}$) that satisfies $\sigma_{\rm v}=\sqrt{\sigma_{\rm th}^2 + \sigma_{\rm kin}^2} = 1670$\,km\,s$^{-1}$ (black curve). 
    The observed line width is indicated as the green region with its statistical uncertainty.
    The red and blue curves indicate the contributions of 
    $\sigma_{\rm th}$ and $\sigma_{\rm kin}$, respectively, 
    for given $V_{\rm rs}$. 
    The dotted curves correspond to the upper and lower limits of the observed line width.
    }\label{fig:v_shock}
\end{figure}

Figure\,\ref{fig:v_shock} shows the relation between $V_{\rm rs}$ and $v_{u,{\rm sh}}$, with which $\sigma_v = 1670_{-170}^{+160}$~km\,s$^{-1}$ is expected (derived using Equations~6 and 10). 
The contributions of the thermal Doppler broadening (red) and the broadening due to the SNR expansion (blue) are also indicated. 
As expected, the former contribution is more significant when $v_{u,{\rm sh}}$ is higher, which is expected for an inward moving reverse shock. 
We find that the observed broadening can be explained when 
$-1000 \lesssim V_{\rm rs}~[{\rm km\,s}^{-1}] \lesssim 3300$, 
indicating that the Fe-rich ejecta in this SNR were shock-heated when the reverse shock was around the turnaround radius. 
This result is reasonable because it is theoretically expected that the reverse shock remains at a standstill for thousands of years in a middle-aged SNR (e.g., \cite{micelotta16}).
If the reverse shock velocity is exactly 0~km\,s$^{-1}$ in observer frame, $v_{u,{\rm sh}} = R_{\rm rs}/t \approx 3700$~km\,s$^{-1}$ is expected. With this upstream velocity, Fe temperature is expected to remain $\gtrsim 1$\,MeV up to $\tau \sim 10^{12}$\,cm$^{-3}$\,s (Figure\,\ref{fig:equil_ej}). Therefore, a significant fraction of the observed line broadening can naturally be attributed to the thermal Doppler broadening.

Although our analysis has successfully provided an estimate for the evolutionary stage of the reverse shock and the thermal properties of the shocked ejecta, it should be emphasized that our analytical model given above is oversimplified. For instance, the geometry of the shocked ejecta must be more complex than we assume, and thus the approximation of the Gaussian-like line profile is likely too simplistic. It is also possible that Fe ions in the shocked ejecta are moderately equilibrated with electrons, and thus the actual thermal Doppler broadening could be slightly smaller. In fact, the Fe Ly$\alpha$ emission shows a narrower width than the Fe He$\alpha$ complex (Table~4), which can be explained if the plasma responsible for the Fe Ly$\alpha$ emission is more equilibrated so that the ion temperature becomes lower. 
We should also note that, if other heavy elements (e.g., Si and S) are also involved in the thermal evolution of the ejecta, the Fe temperature drops more quickly than is predicted in Figure\,\ref{fig:equil_ej} due to the Coulomb interactions among the heavy elements. 
If the initial Fe temperature is substantially higher than the current temperature, the required upstream velocity becomes higher than our estimates in Figure\,\ref{fig:v_shock}.
To constrain the physical quantities more precisely, detailed calculations based on hydrodynamical simulations need to be employed.

\section{Conclusions}

We have presented analysis of the XRISM first-light observation data of N132D, the X-ray brighest SNR in the LMC. 
The excellent performance of the X-ray microcalorimeter Resolve has enabled us to perform high resolution spectroscopy of this SNR in the 1.6--10-keV band, for the first time. 
Our analysis has revealed that the K-shell emission lines of Si and S, whose origin is thought to be the swept-up ISM, are mildly broadened with $\sigma_v \sim 450$\,km\,s$^{-1}$. 
Under an assumption of a nearly symmetrically expanding ISM shell, the radial component of the blast wave velocity is estimated to be $\sim 1200$\,km\,s$^{-1}$, lower than the recent proper motion measurement with Chandra. 
On the other hand, the Fe H$\alpha$ complex emission is substantially broadened with $\sigma_v \sim 1670$\,km\,s$^{-1}$.
If this emission originates from the ejecta, the observed line width can be explained through a combination of the thermal Doppler broadening due to the high ion temperature in the non-equilibrium plasma and kinematic Doppler effect due to the expansion of the shocked ejecta. We have also provided an estimate for the evolutionary stage of the SNR, putting a constraint on the reverse shock velocity (in observer frame) to be $-1000 \lesssim V_{\rm rs}~[{\rm km\,s}^{-1}] \lesssim 3300$, the value at the time when the bulk of the Fe ejecta were shock-heated. 
If $V_{\rm rs} \approx 0$\,km\,s$^{-1}$, which is reasonably expected from the theoretical point of view, the current Fe temperature is estimated to be $\gtrsim 1$\,MeV. 
The redshift observed in the Si, S, and Fe He$\alpha$ lines is equivalent to the radial velocity of the ISM surrounding N132D ($\sim 250$\,km\,s$^{-1}$), whereas that observed in Fe Ly$\alpha$ indicates a substantially larger radial velocity of 890\,km\,s$^{-1}$, about 600\,km\,s$^{-1}$ greater than the radial velocity of the local ISM in the LMC. Also our imaging analysis revealed that the Fe Ly$\alpha$ emission is concentrated around the SNR center. 
These results suggest that the Fe Ly$\alpha$ emission originates from hot Fe ejecta that are present only at the far side of the SNR, distinct from the component responsible for the Fe He$\alpha$ emission. 

This work represents the first paper to be published by the XRISM Collaboration. The results presented here are uniquely obtained using the spectral power of the Resolve combined with the imaging of Xtend, offering only a modest glimpse of the power of this observatory that will revolutionize our understanding of the Universe.


\begin{ack}
The XRISM team acknowledges the hundreds, likely thousands, of scientists and engineers in Japan, the United States, Europe, and Canada who contributed to not only this mission, but to all predecessors that came before. This mission is a testament to the long-standing collaborations between the countries and space agencies involved. 
The authors deeply thank Prof.\ Kiyoshi Hayashida, who passed away on October 2, 2021, for his significant contribution to the project and whole X-ray astronomy. 
HY\ is thankful to Dr.\ Anne Decourchelle for her helpful comments on this manuscript and to Dr.\ Daniel Patnaude for discussion about interpretation of the observational results.
We also thank the anonymous referee for an insightful and constructive review that improved this work.

This work was supported by JSPS KAKENHI grant numbers JP22H00158, JP22H01268, JP22K03624, JP23H04899, JP21K13963, JP24K00638, JP24K17105, JP21K13958, JP21H01095, JP23K20850, JP24H00253, JP21K03615, JP24K00677, JP20K14491, JP23H00151, JP19K21884, JP20H01947, JP20KK0071, JP23K20239, JP24K00672, JP24K17104, JP24K17093, JP20K04009, JP21H04493, JP20H01946, JP23K13154, JP19K14762, JP20H05857, JP23K03459, and  JP22KJ1047, and NASA grant numbers 80NSSC20K0733, 80NSSC18K0978, 80NSSC20K0883, 80NSSC20K0737, 80NSSC24K0678, 80NSSC18K1684, and 80NNSC22K1922. 
LC acknowledges support from NSF award 2205918. CD acknowledges support from STFC through grant ST/T000244/1. LG acknowledges financial support from Canadian Space Agency grant 18XARMSTMA. AT and the present research are in part supported by the Kagoshima University postdoctoral research program (KU-DREAM). SY acknowledges support by the RIKEN SPDR Program. IZ acknowledges partial support from the Alfred P. Sloan Foundation through the Sloan Research Fellowship. 
Part of this work was performed under the auspices of the U.S. Department of Energy by Lawrence Livermore National Laboratory under Contract DE-AC52-07NA27344. The material is based upon work supported by NASA under award number 80GSFC21M0002. 
This work was supported by the JSPS Core-to-Core Program, JPJSCCA20220002. The material is based on work supported by the Strategic Research Center of Saitama University.

\end{ack}


\appendix 
\section{Resolve Gain calibration}

The Resolve detector gain and energy assignment requires correction of the time dependent gain on-orbit. The Resolve calorimeter detectors are thermal detectors and they thus have both a bolometric and transient response. The bolometric response reflects the thermal radiation environment within the Resolve instrument and impacts the gain of the transient response to X-rays. In addition, the detector gain is impacted by the detector heat sink temperature which is regulated to better than 1 $\mu$K on orbit, and the temperature of the amplifier and control electronics that are affected by spacecraft orientation. To compensate for these effects, the Resolve detector gain is tracked as a function of time using on-board calibration sources. The time dependent energy scale is then reconstructed as a function of time by interpolating a family of temperature dependent gain curves measured during ground calibration \citep{porter16}. To measure the gain, a set of $^{55}$Fe radioactive sources on the filter wheel are periodically rotated into the aperture of the instrument. The Mn K$\alpha$ X-rays from the radioactive sources are fit using an empirically measured core line-shape \citep{Hoelzer97} for each fiducial interval. For N132D, gain fiducials were measured every orbit for $\approx 30$ minutes during earth occultation yielding 500--600 counts in the Mn K$\alpha$ line complex giving a statistical uncertainty in the energy scale of between 0.15--0.2\,eV at 5.9\,keV for each fiducial interval. N132D was observed early in the commissioning phase for the XRISM mission, before the final time dependent methodology was adopted. In later observations, the time dependent gain fiducial measurements were optimized since far fewer fiducial measurements are needed to track the detector gain. The N132D observations discussed here included 72 gain fiducial measurements, far in excess of what is needed to track the gain with equivalent precision. Later observations of equivalent length would have included only 15 fiducial measurements.

\begin{figure}[t]
 \begin{center}
  \includegraphics[width=7.8cm]{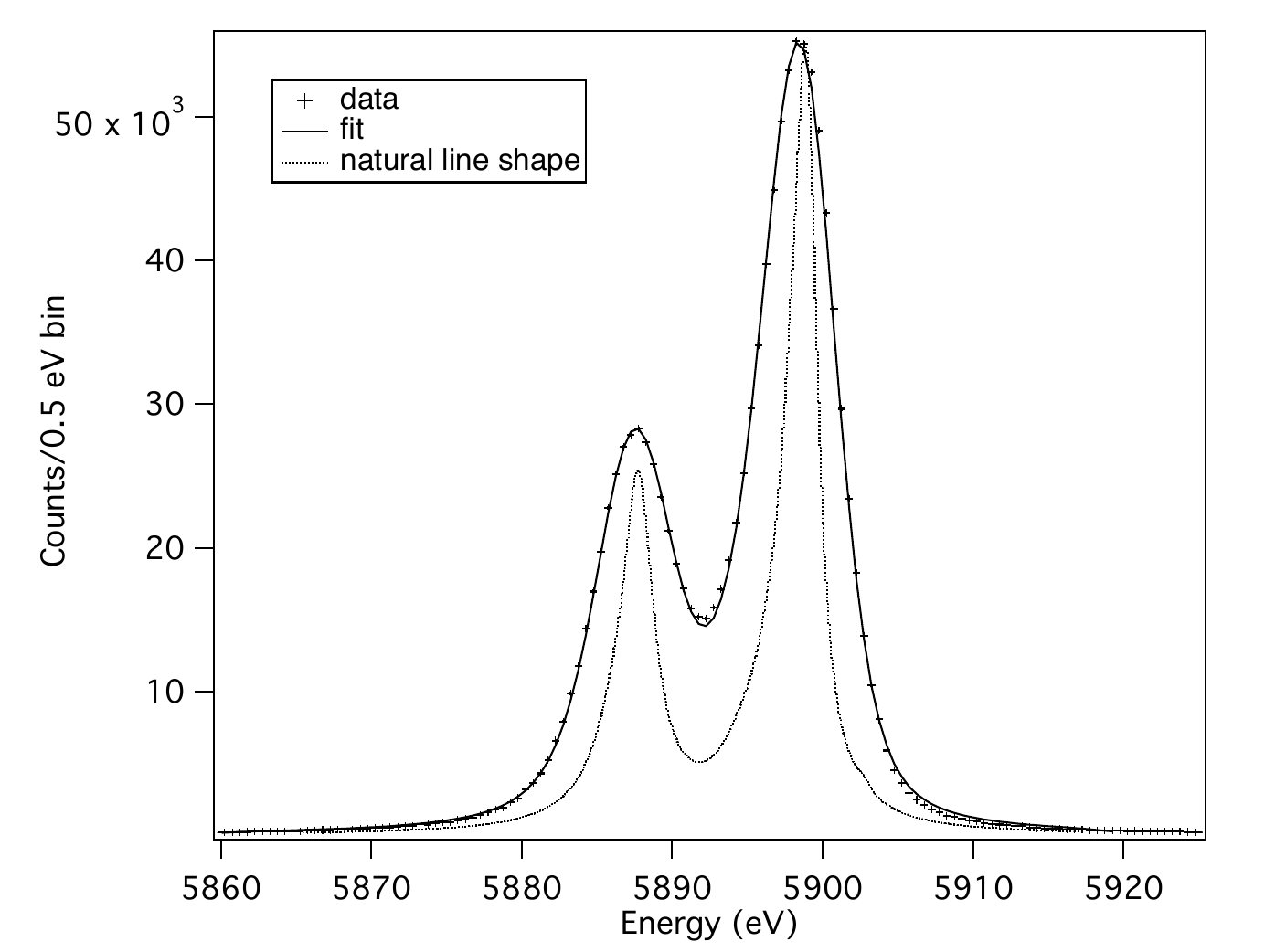} 
 \end{center}
    \caption{The Resolve spectrum of Mn K$\alpha$ X-rays from $^{55}$Fe radioactive sources on the filter wheel. The spectrum is a composite of the pixels in the main array and measured during fiducial intervals once per orbit during earth occultation. A fit using the standard Gaussian instrumental function yields an energy resolution of  4.43 eV (FWHM), and an energy scale error of 0.04 eV after energy scale reconstruction.
    }\label{fig:array_fiducial}
\end{figure}

\begin{figure}[t]
 \begin{center}
  \includegraphics[width=7.8cm]{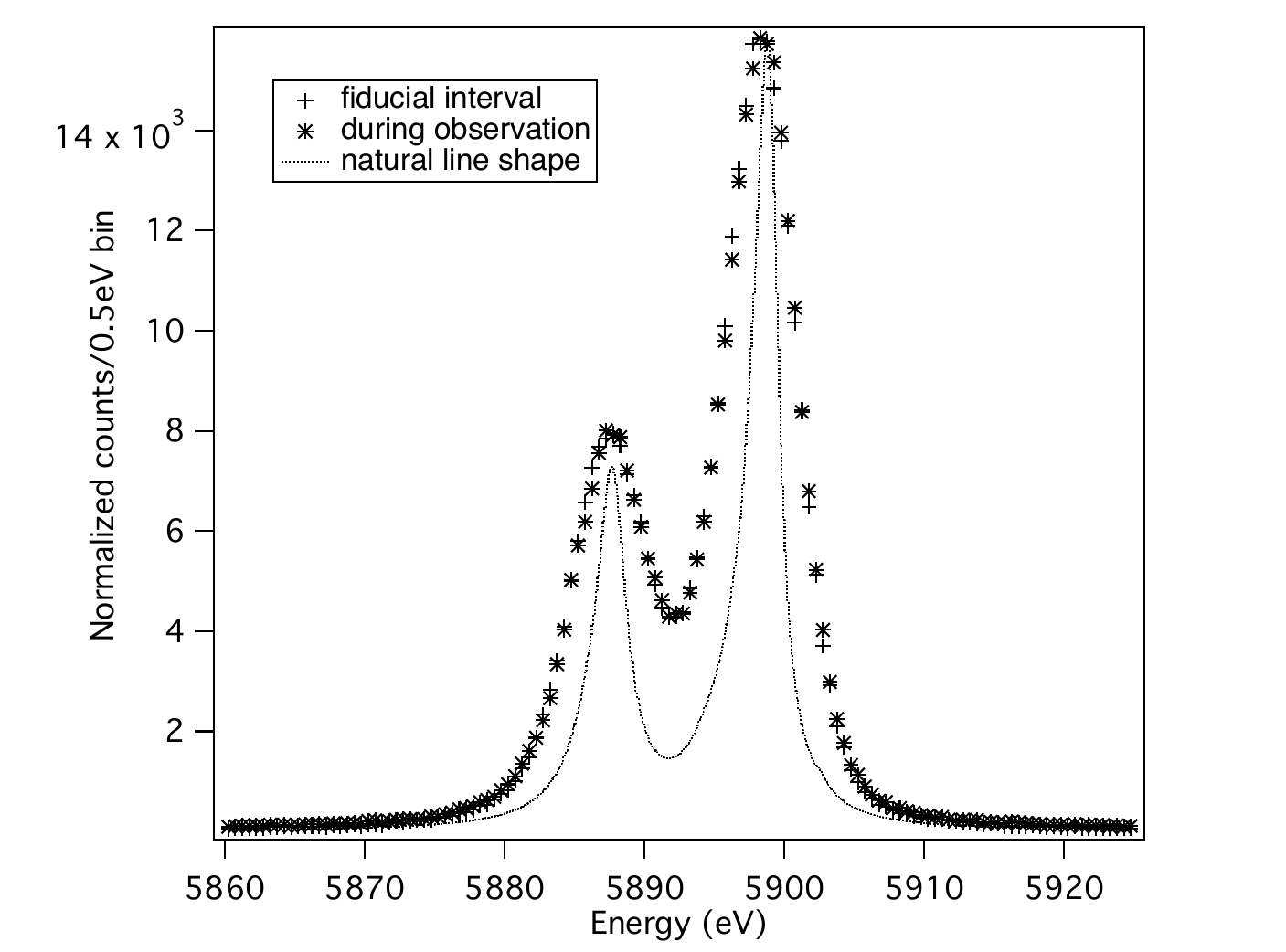} 
 \end{center}
    \caption{Resolve calibration pixel spectrum of Mn K$\alpha$ X-rays from a pencil beam $^{55}$Fe radioactive sources that continuously illuminates the calibration pixel. Two data sets are shown, one during the same fiducial intervals used to track the gain of the main array, and the other just during the observation, not including the fiducial intervals. In both cases, the energy scale reconstruction is the same as that used for the main array and the comparison demonstrates the efficacy of the energy scale reconstruction for this observation. Fits using a Gaussian instrumental function yield energy resolutions of 4.42 and 4.39 eV (FWHM), and energy scale errors of 0.04 and 0.11 eV at 6 keV respectively.
    }\label{fig:calpix}
\end{figure}

In order to assess the energy scale reconstruction and the detector energy resolution during the observation, we perform several additional checks using high-resolution primary (Hp) grade events. The first is to fit the fiducial Mn K$\alpha$ complex per-pixel and as an array composite after energy scale reconstruction. Figure\,\ref{fig:array_fiducial} shows the array composite Mn K$\alpha$ complex, the underlying natural line shape \citep{Hoelzer97}, and a fit using a Gaussian instrumental function. The fit gives an energy resolution of 4.43 eV (FWHM), and an energy scale error of 0.04 eV, both of which are within 0.1 eV of the standard on-orbit performance. Additionally, we use a calibration pixel to verify the performance of the instrument during the main observation outside of the gain fiducials. Resolve includes a standard pixel, identical to the main array, but just outside the field of view. The calibration pixel is continuously illuminated with a pencil beam $^{55}$Fe radioactive source allowing the detector gain and energy resolution to be continuously tracked. Using the calibration pixel, we compare the energy scale reconstruction and energy resolution during the same fiducial intervals as the main array and also just during the observation but using the same energy scale reconstruction method as the main array. These two data sets are shown in Figure\,\ref{fig:calpix}. Fits using the standard Gaussian instrumental function yield an energy resolution of 4.42 eV (FWHM) and an energy scale error of 0.04 eV during the fiducial intervals, and a resolution of 4.39 eV (FWHM) and an energy scale error of 0.11 eV during the N132D observations. On-orbit measurements using Cr and Cu fluorescent sources and a Si instrumental line, give array composite systematic uncertainties in the energy scale of $< 0.2$\,eV in the 5.4--8.0\,keV band and at most 1.3\,eV at the low energy edge of the band at 1.75\,keV. On-orbit energy scale measurements and analysis are on-going and we expect that these uncertainties will be reduced in the future with a goal of $< 0.1$\,eV across the Resolve bandpass.

\section{Resolve Non X-ray Background}

We developed a model for the Resolve instrumental background (or non-X-ray background, NXB) based on $\sim$\,7~month data accumulated during periods of Earth eclipse, supplemented by the Hitomi SXS NXB \citep{kilbourne18} and ``blank-sky" data. Starting with the Resolve eclipse data, we applied standard screening, identical to that applied to the on-source data, and produced a spectrum composed by aggregating data from periods of different cut-off rigidity according to the weighting found in the N132D observations. The NXB level is $< 10^{-3}$ s$^{-1}$ keV$^{-1}$ for the entire array over the energy range of interest.

We developed a model for the NXB continuum from this early Resolve NXB data set, but we did not fit the instrumental lines due to distortions and shifts introduced by the initial per-pixel energy scales of the many eclipse segments. Instead, we turned to the Hitomi SXS NXB, determined the strengths of the detected instrument lines, and compared these to the line strengths in 226~ks of Resolve blank-sky data with better aligned pixel energy scales than the preliminary Resolve NXB data set.

The amplitude of the Au L$\alpha_1$ line is consistent between the two data sets, but the Mn K$\alpha$ line is significantly weaker in the Resolve data, as expected from ground data.  This feature is the result of scattered X-rays from the collimated $^{55}$Fe source pointed at the dedicated calibration pixel, the design of which was modified for Resolve. We determined that the Mn K$\alpha$ line in the Resolve blank-sky data is consistent with that determined in a high-statistics ground measurement, adjusted for the half-life of $^{55}$Fe. The statistics of the other lines were not adequate to inform the model, thus we used the SXS line strengths for all the instrumental lines except Mn K$\alpha$. We approximated the following 12 lines by Gaussians: Al K$\alpha$, Au M$\alpha_1$, Mn K$\alpha_1$, Mn K$\alpha_2$, Ni K$\alpha_1$, Ni K$\alpha_2$, Cu K$\alpha_1$, Cu K$\alpha_2$, Au L$\alpha_1$, Au L$\alpha_2$, Au L$\beta_1$, and  Au L$\beta_2$. Although better line profiles are known for most of these lines, the statistics of the observation do not warrant a more accurate specification in the model. We separately specified the K$\alpha$ doublets to capture the widths of these profiles, and fixed their normalizations at 2:1. When applying the NXB model to the data of N132D, we adjusted the normalizations of the Mn K$\alpha$, Ni K$\alpha$, and Au L$\alpha_1$ lines, so that their intensities match the observation.

\section{Line Profile for Symmetrically Expanding Shell and Sphere}

\begin{figure*}[t]
 \begin{center}
  \includegraphics[width=15.2cm]{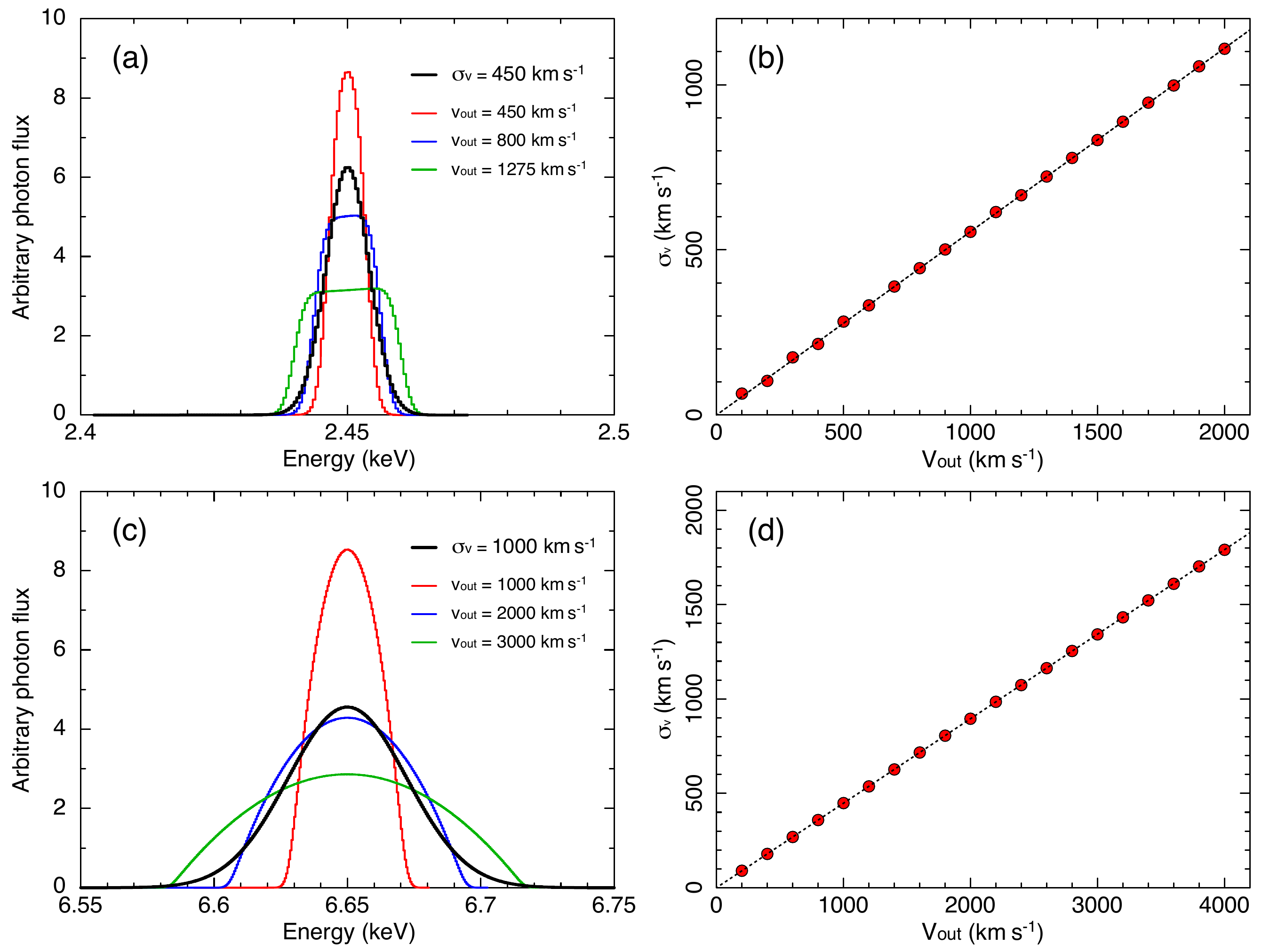} 
 \end{center}
    \caption{(a) Profiles of a single emission line at 2.45\,keV, expected for the expanding shell model with 
    $v_{\rm out} = 450$~km\,s$^{-1}$ (red), 800~km\,s$^{-1}$ (blue), and 1275~km\,s$^{-1}$ (green), compared with a velocity-broadened Gaussian with $\sigma_v = 450$~km\,s$^{-1}$ (black). \ 
    (b) The relation between $\sigma_v$ and $v_{\rm out}$ for the expanding shell model. See text for details. \ 
    (c) Profiles of a single emission line at 6.65\,keV, expected for the expanding uniform sphere model with 
    $v_{\rm out} = 1000$~km\,s$^{-1}$ (red), 2000~km\,s$^{-1}$ (blue), and 3000~km\,s$^{-1}$ (green), compared with a velocity-broadened Gaussian with $\sigma_v = 1000$~km\,s$^{-1}$ (black). \
    (d) Same as Panel (b), but for the uniform sphere model. \
    The spectra in Panel (a) and (c) are convolved with the Resolve RMF and ARF. 
    }\label{fig:profile}
\end{figure*}

In our spectral analysis, a Gaussian velocity broadening (implemented in the \texttt{bvrnei} model in XSPEC) is assumed to fit the broadened emission lines. However, broadening due to the SNR expansion depends on the radial velocity distribution of the shocked plasma, being not necessarily Gaussian-like. Here we investigate line profiles expected for symmetrically expanding shell and sphere, and compare them with the Gaussian profiles. 

First we consider a spherically symmetric shell with a thickness of $R_{\rm out}/12$, where $R_{\rm out}$ is the outermost radius (and thus the inner radius is $11R_{\rm out}/12$), typically expected for the swept-up ISM of SNRs in the Sedov phase. The expansion velocity at the radius $r$ is assumed to be $v = v_{\rm out} \cdot (r/R_{\rm out})$, where $v_{\rm out}$ is the velocity at the radius $R_{\rm out}$. We also assume uniform plasma density within the shell. 
Figure~\ref{fig:profile}a shows the expected profiles of a single emission line at 2.45\,keV (corresponding to the S He$\alpha$ emission) for $v_{\rm out} = 450$~km\,s$^{-1}$ (red), 800~km\,s$^{-1}$ (blue), and 1275~km\,s$^{-1}$ (green), compared with a velocity-broadened Gaussian with $\sigma_v = 450$~km\,s$^{-1}$ (black). 
The spectra are convolved with the Resolve RMF and ARF generated in Section~2.  
We find that a similar line width is expected in the expanding shell model with $v_{\rm out} = 800$~km\,s$^{-1}$ and in the Gaussian with $\sigma_v = 450$~km\,s$^{-1}$. 
We then generate mock Resolve spectra for various $v_{\rm out}$ values ranging from 100 to 2000~km\,s$^{-1}$ and fit them with a Gaussian model. 
Figure~\ref{fig:profile}b shows the relation between the best-fit $\sigma_v$ values and input $v_{\rm out}$, approximated by a linear function of $\sigma_v \approx 0.55 \, v_{\rm out}$. 
Note that the expanding shell model predicts a characteristic top-flat profile. For this reason, a Gaussian function is not a good approximation, especially when the outermost expansion velocity is high. 

Next we consider a uniform density sphere. Again, the expansion velocity is assumed to be proportional to the radius, $v = v_{\rm out} \cdot (r/R_{\rm out})$. 
Figure~\ref{fig:profile}c shows the expected profiles of a single emission line at 6.65\,keV (corresponding to the Fe He$\alpha$ emission) for $v_{\rm out} = 1000$~km\,s$^{-1}$ (red), 2000~km\,s$^{-1}$ (blue), and 3000~km\,s$^{-1}$ (green), compared with a velocity-broadened Gaussian with $\sigma_v = 1000$~km\,s$^{-1}$ (black). 
Among the three, the second profile gives the best approximation to the Gaussian. 
Figure~\ref{fig:profile}d shows the relation between $\sigma_v$ and $v_{\rm out}$ for the uniform sphere model, derived similarly to Figure~\ref{fig:profile}b.
We find that the relation is approximated by a linear function of $\sigma_v \approx 0.45 \, v_{\rm out}$. 

We make similar investigations for a spherically symmetric shell with different thickness of $R_{\rm out}/12 < l_{\rm shell} < R_{\rm out}$, and find that the relation of $\sigma_v \approx \alpha \cdot v_{\rm out}$ is obtained with $0.45 \lesssim \alpha \lesssim 0.55$. 
We also find that their is no substantial dependence on the photon energy in this relation.
Considering the complexity in actual velocity distribution, we simply assume $\sigma_{\rm kin} \approx 0.5 \, v_{\rm out}$ in Section~4.2.


\begin{thebibliography}{}

\bibitem[Acero et al.(2016)]{acero16} Acero, F., Ackermann, M., Ajello, M., et al.\ 2016, \apjs, 224, 8

\bibitem[Ackermann et al.(2016)]{ackermann16} Ackermann, M., Albert, A., Atwood, W.~B., et al.\ 2016, \aap, 586, A71

\bibitem[Arnaud(1996)]{arnaud96} Arnaud, K.~A.\ 1996, Astronomical Data Analysis Software and Systems V, 101, 17

\bibitem[Bamba et al.(2018)]{bamba18} Bamba, A., Ohira, Y., Yamazaki, R., et al.\ 2018, \apj, 854, 71

\bibitem[Banovetz et al.(2023)]{banovetz23} Banovetz, J., et al. 2023, ApJ, 948, 33

\bibitem[Behar et al.(2001)]{behar01} Behar, E., et al. 2001, A\&A, 365, 242

\bibitem[Blair et al.(2000)]{blair00} Blair, W.~P., Morse, J.~A., Raymond, J.~C., et al.\ 2000, \apj, 537, 667

\bibitem[Blondin et al.(2003)]{blondin03} Blondin, J.~M., Mezzacappa, A., \& DeMarino, C.\ 2003, \apj, 584, 971

\bibitem[Boggs et al.(2015)]{boggs15} Boggs, S.~E., Harrison, F.~A., Miyasaka, H., et al.\ 2015, Science, 348, 670

\bibitem[Borkowski et al.(2007)]{borkowski07} Borkowski, K.J., Hendrick, S.P., \& Reynolds, S.P. 2007, ApJ, 671, 45

\bibitem[Borkowski et al.(2020)]{borkowski20} Borkowski, K.~J., Miltich, W., \& Reynolds, S.~P.\ 2020, \apjl, 905, L19

\bibitem[Cash(1979)]{cash79} Cash, W.\ 1979, \apj, 228, 939

\bibitem[Danziger \& Dennefeld(1976)]{danziger76} Danziger, I.~J. \& Dennefeld, M.\ 1976, \apj, 207, 394

\bibitem[DeLaney et al.(2010)]{delaney10} DeLaney, T., Rudnick, L., Stage, M.~D., et al.\ 2010, \apj, 725, 2038

\bibitem[Dickel \& Milne(1995)]{dickel95} Dickel, J.R. \& Milne, D.K. 1995, A\&A, 109, 200

\bibitem[Dickey \& Lockman(1990)]{dickey90} Dickey, J.~M. \& Lockman, F.~J.\ 1990, \araa, 28, 215

\bibitem[Dopita \& Tuohy(1984)]{dopita84} Dopita, M.~A. \& Tuohy, I.~R.\ 1984, \apj, 282, 135

\bibitem[Dopita et al.(2019)]{dopita19} Dopita, M.~A., Seitenzahl, I.~R., Sutherland, R.~S., et al.\ 2019, \aj, 157, 50

\bibitem[Favata et al.(1997)]{favata97} Favata, F., Vink, J., Parmar, A.~N., et al.\ 1997, \aap, 324, L45

\bibitem[Ghavamian et al.(2007)]{ghavamian07} Ghavamian, P., Laming, J.~M., \& Rakowski, C.~E.\ 2007, \apjl, 654, L69

\bibitem[Grefenstette et al.(2014)]{grefenstette14} Grefenstette, B.~W., Harrison, F.~A., Boggs, S.~E., et al.\ 2014, \nat, 506, 339

\bibitem[H.E.S.S. Collaboration(2021)]{hess21} H.E.S.S. Collaboration 2021, A\&A, 655, 7

\bibitem[Hitomi Collaboration(2018)]{hitomi18} Hitomi Collaboration 2018, PASJ, 70, 16

\bibitem[Hoelzer et al.(1997)]{Hoelzer97} Hoelzer, G., Fritsch, M., Deutsch, M. et al.\ 2016, Phys.\ Rev.\ A, 56, 4554 

\bibitem[Hughes(1987)]{hughes87} Hughes, J.P. 1987, ApJ, 314, 103

\bibitem[Hughes et al.(1998)]{hughes98} Hughes, J.P., Hayashi, I., \& Koyama, K. 1998, ApJ, 505, 732

\bibitem[Hwang et al.(1993)]{hwang93} Hwang, U., Hughes, J.P., Canizares, C.R., \& Markert, T.H. 1993, ApJ, 414, 219

\bibitem[Hwang \& Laming(2012)]{hwang12} Hwang, U. \& Laming, J.~M.\ 2012, \apj, 746, 130

\bibitem[Ishisaki et al.(2022)]{ishisaki22} Ishisaki, Y., Kelley, R.~L., Awaki, H., et al.\ 2022, \procspie, 12181, 121811S

\bibitem[Janka et al.(2016)]{janka16} Janka, H.-T., Melson, T., \& Summa, A.\ 2016, Annual Review of Nuclear and Particle Science, 66, 341

\bibitem[Kaastra \& Bleeker(2016)]{kaastra16} Kaastra, J.~S. \& Bleeker, J.~A.~M.\ 2016, \aap, 587, A151

\bibitem[Kilbourne et al.(2018)] {kilbourne18} Kilbourne, C.A., et al. 2018, PASJ, 70, 18

\bibitem[Lasker(1978)]{lasker78} Lasker, B.M. 1978, ApJ, 223, 109

\bibitem[Lasker(1980)]{lasker80} Lasker, B.~M.\ 1980, \apj, 237, 765

\bibitem[Law et al.(2020)]{law20} Law, C.~J., Milisavljevic, D., Patnaude, D.~J., et al.\ 2020, \apj, 894, 73

\bibitem[Long \& Helfand(1979)]{long79} Long, K.S. \& Helfand, D.J. 1979, ApJ, 234, 77

\bibitem[Maggi et al.(2016)]{maggi16} Maggi, P., Haberl, F., Kavanagh, P.~J., et al.\ 2016, \aap, 585, A162

\bibitem[Mathewson et al.(1983)]{mathewson83} Mathewson, D.~S., Ford, V.~L., Dopita, M.~A., et al.\ 1983, \apjs, 51, 345

\bibitem[Micelotta et al.(2016)]{micelotta16} Micelotta, E.~R., Dwek, E., \& Slavin, J.~D.\ 2016, \aap, 590, A65

\bibitem[Midooka et al.(2020)]{midooka20} Midooka, T., et al. 2020, Proc SPIE, 11444, 114445C

\bibitem[Milisavljevic \& Fesen(2013)]{milisavljevic13} Milisavljevic, D. \& Fesen, R.~A.\ 2013, \apj, 772, 134

\bibitem[Mori et al.(2022)]{mori22} Mori, K., Tomida, H., Nakajima, H., et al.\ 2022, \procspie, 12181, 121811T. doi:10.1117/12.2626894

\bibitem[Morse et al.(1995)]{morse95} Morse, J.A., Winkler, P.F., \& Kirshner, R.P. 1995, AJ, 109, 2104

\bibitem[Plucinsky et al.(2024)]{plucinsky24} Plucinsky, P., Long, X., Kashyap, V., et al.\ 2024, American Astronomical Society Meeting

\bibitem[Pietrzy{\'n}ski et al.(2013)]{pietrzynski13} Pietrzy{\'n}ski, G., Graczyk, D., Gieren, W., et al.\ 2013, \nat, 495, 76

\bibitem[Podsiadlowski(2017)]{podsiadlowski17} Podsiadlowski, P.\ 2017, Handbook of Supernovae, 635

\bibitem[Porter et al.(2016)]{porter16} Porter, F. S., Chiao, M. P., Eckart, M.E. et al.\ 2016, J. Low Temp Phys, 184, 498

\bibitem[Ravi et al.(2024)]{ravi24} Ravi, A.~P., Park, S., Zhekov, S.~A., et al.\ 2024, \apj, 966, 147

\bibitem[Rho et al.(2023)]{rho23} Rho, J., Ravi, A.~P., Slavin, J.~D., et al.\ 2023, \apj, 949, 74

\bibitem[Russell \& Dopita(1992)]{russell92} Russell, S.~C. \& Dopita, M.~A.\ 1992, \apj, 384, 508

\bibitem[Sano et al.(2020)]{sano20} Sano, H., Plucinsky, P.~P., Bamba, A., et al.\ 2020, \apj, 902, 53

\bibitem[Sharda et al.(2020)]{sharda20} Sharda, P., Gaetz, T.~J., Kashyap, V.~L., et al.\ 2020, \apj, 894, 145

\bibitem[Smith(2017)]{smith17} Smith, N.\ 2017, Handbook of Supernovae, 403

\bibitem[Sutherland \& Dopita(1995)]{sutherland95} Sutherland, R.S. \& Dopita, M.A. 1995, ApJ, 439, 365

\bibitem[Suzuki et al.(2020)]{suzuki20} Suzuki, H., Yamaguchi, H., Ishida, M., et al.\ 2020, \apj, 900, 39

\bibitem[Tashiro et al.(2020)]{tashiro20} Tashiro, M., Maejima, H., Toda, K., et al.\ 2020, \procspie, 11444, 1144422

\bibitem[Truelove \& McKee(1999)]{truelove99} Truelove, J.~K. \& McKee, C.~F.\ 1999, \apjs, 120, 299

\bibitem[Tsuchioka et al.(2021)]{tsuchioka21} Tsuchioka, T., Uchiyama, Y., Higurashi, R., et al.\ 2021, \apj, 912, 131

\bibitem[Vink et al.(2015)]{vink15} Vink, J., Broersen, S., Bykov, A., et al.\ 2015, \aap, 579, A13

\bibitem[Vink et al.(2022)]{vink22} Vink, J., Patnaude, D.~J., \& Castro, D.\ 2022, \apj, 929, 57. doi:10.3847/1538-4357/ac590f

\bibitem[Vogt \& Dopita(2011)]{vogt11} Vogt, F. \& Dopita, M.~A.\ 2011, \apss, 331, 521

\bibitem[Westerlund \& Mathewson(1966)]{westerlund66} Westerlund, B.E. \& Mathewson, D.S. 1966, MNRAS, 131, 371

\bibitem[Williams et al.(2006)]{williams06} Williams, B.J., et al. 2006, ApJ, 652, 33

\bibitem[Yamaguchi et al.(2014a)]{yamaguchi14a} Yamaguchi, H., Badenes, C., Petre, R., et al.\ 2014, \apjl, 785, L27

\bibitem[Yamaguchi et al.(2014b)]{yamaguchi14b} Yamaguchi, H., Eriksen, K.~A., Badenes, C., et al.\ 2014, \apj, 780, 136.

\end{thebibliography}
\end{document}